\documentclass[iop,numberedappendix,twocolappendix,appendixfloats]{emulateapjKA}
\usepackage{graphicx,amsmath, amsfonts, amssymb,aas_macros,times,natbib,epstopdf}
\usepackage{subfigure, xspace, comment}
\usepackage[usenames,dvipsnames,svgnames,x11names,table]{xcolor}
\usepackage{soul}
\usepackage[export]{adjustbox}
\usepackage{soul}
\usepackage{threeparttable}
\usepackage[singlelinecheck=false]{caption}
\usepackage[bookmarks=false]{hyperref}
\hypersetup{
  colorlinks= true, 
  urlcolor  = blue, 
  filecolor=black,
  runcolor=blue,
  linkcolor = RubineRed, 
  citecolor = blue 
} 
\newcommand{\angstrom}{\textup{\AA}} 
\newcommand{\GALEX}{\textit{GALEX}\xspace}

\shorttitle{UV colors of SPOGs}
\shortauthors{Ardila et al.}

\begin{document}


\title{Shocked POststarburst Galaxy Survey. III. The Ultraviolet Properties of SPOGs}
\author{Felipe Ardila$^{1}$, Katherine Alatalo$^{2,3}$\altaffilmark{$\dagger$}, Lauranne Lanz$^{4}$, Philip N. Appleton$^{5}$, Rachael L. Beaton$^{3,6}$\altaffilmark{$\dagger$}\altaffilmark{$*$}, Theodoros Bitsakis$^{7}$, Sabrina L. Cales$^{8}$, Jes{\'u}s Falc{\'o}n-Barroso$^{9,10}$, Lisa J. Kewley$^{11}$, Anne M. Medling$^{12,13}$\altaffilmark{$\dagger$}, John S. Mulchaey$^{3}$, Kristina Nyland$^{14}$, Jeffrey A. Rich$^{3}$, \& C. Meg Urry$^{8}$}

\affil{
$^{1}$ Department of Astronomy and Astrophysics, University of California, Santa Cruz, CA 95064, USA\\
$^{2}$Space Telescope Science Institute, 3700 San Martin Dr., Baltimore, MD 21218, USA\\
$^{3}$The Observatories of the Carnegie Institution for Science, 813 Santa Barbara St., Pasadena, CA 91101, USA\\
$^{4}$Dartmouth College, 6127 Wilder Laboratory, Hanover, NH 03755, USA\\
$^{5}$Infrared Processing and Analysis Center, California Institute of Technology, Pasadena, CA 91125, USA\\
$^{6}$ Department of Astrophysical Sciences, Princeton University, Princeton, NJ 08544, USA\\
$^{7}$Instituto de Radioastronom{\'i}a y Astrof{\'i}sica, Universidad Nacional Aut{\'o}noma de M{\'e}xico, C.P. 58190, Morelia, Mexico\\
$^{8}$Yale Center for Astronomy and Astrophysics, Physics Department, Yale University, New Haven, CT 06511 USA\\
$^{9}$Instituto de Astrof{\'i}sica de Canarias, E-38205 La Laguna, Tenerife, Spain\\
$^{10}$Departamento de Astrof{\'i}sica, Universidad de La Laguna (ULL), E-38200 La Laguna, Tenerife, Spain\\
$^{11}$Research School of Astronomy and Astrophysics, Australian National University, Cotter Rd., Weston ACT 2611, Australia\\
$^{12}$Cahill Center for Astronomy and Astrophysics, California Institute of Technology, MS 249-17 Pasadena, CA 91125, USA\\
$^{13}$Research School for Astronomy \& Astrophysics, Australian National University, Canberra, ACT 2611, Australia\\
$^{14}$National Radio Astronomy Observatory, 520 Edgemont Road, Charlottesville, VA 22903, USA}

\altaffiltext{$\dagger$}{Hubble Fellow}
\altaffiltext{$*$}{Carnegie-Princeton Fellow}
\email{fardila@ucsc.edu}

\begin{abstract}
The Shocked POststarburst Galaxy Survey (SPOGS) aims to identify galaxies in the transitional phase between actively star-forming and quiescence with nebular lines that are excited from shocks rather than star formation processes. We explored the ultraviolet (UV) properties of objects with near-ultraviolet (NUV) and far-ultraviolet (FUV) photometry from archival \textit{GALEX} data; 444 objects were detected in both bands, 365 in only NUV, and 24 in only FUV, for a total of 833 observed objects. We compared SPOGs to samples of Star-forming galaxies (SFs), Quiescent galaxies (Qs), classical E+A post-starburst galaxies, active galactic nuclei (AGN) host galaxies, and interacting galaxies. We found that SPOGs have a larger range in their FUV--NUV and NUV--\textit{r} colors compared to most of the other samples, although all of our comparison samples occupied color space inside of the SPOGs region. Based on their UV colors, SPOGs are a heterogeneous group, possibly made up of a mixture of SFs, Qs, and/or AGN. Using Gaussian mixture models, we are able to recreate the distribution of FUV--NUV colors of SPOGs and E+A galaxies with different combinations of SFs, Qs, and AGN. We find that the UV colors of SPOGs require a $>$60\% contribution from SFs, with either Qs or AGN representing the remaining contribution, while UV colors of E+A galaxies required a significantly lower fraction of SFs, supporting the idea  that SPOGs are at an earlier point in their transition from quiescent to star-forming than E+A galaxies.

\end{abstract}

\keywords{galaxies: evolution --- galaxies: star formation -- galaxies: stellar content -- ultraviolet: galaxies}

\section{Introduction}\label{sec:intro}

There is an observed bimodality in morphology \citep{hubble1926,strateva+01,baldry+04}, color \citep{baade1958,tinsley1978}, star formation rates (SFRs), and gas fractions in the population of present-day galaxies. In color-magnitude space, this bimodality is seen as a red sequence and a blue cloud \citep{strateva+01,baldry+04}. The red sequence includes ``early-type'' galaxies (ETGs), which are relatively gas-poor, have redder optical colors, ellipsoidal morphologies, and typically quenched star formation. Blue cloud galaxies, on the other hand, are usually ``late-type'' galaxies (LTGs), which are gas-rich, have bluer colors, flattened disk morphologies, and are actively forming stars.

From z = 1 $\rightarrow$ 0, the total mass of blue cloud galaxies has remained roughly constant \citep{noeske+2007}, while the red sequence has doubled in mass \citep{bell+2012}. This suggests that once its star formation is quenched, a blue cloud galaxy migrates to the red sequence \citep{harker+2006}. The reverse migration (from red sequence to blue cloud) does not commonly occur \citep{young+2014}, except in gas-rich mergers or other such extreme events \citep{kannappan+2009}. Few galaxies are seen in the intermediate ``green valley'' space of color-magnitude diagrams. It has been suggested that this lack is due to the rapid timescales of the transition from blue cloud to red sequence \citep{faber+2007}. While there is a correlation between galaxy morphology and star formation rate, it does not necessarily imply causation. Analyses with the Sloan Digital Sky Survey (SDSS), \textit{Galaxy Evolution Explorer} \citep[\GALEX;][]{martin+2005}, and Galaxy Zoo \citep{lintott+2008} data have shown that while morphological ETGs do transition rapidly through the green valley, LTGs do not; instead they retain their morphologies as specific star formation rates decline very slowly \citep{schawinski+2014}.

These differing rates of transition suggest that there may be several transitional states within the green valley. It has been proposed that star-forming galaxies could transition as a result of the cosmic supply of gas being shut off (e.g., at a critical halo mass), and depletion of the remaining gas (via secular or external means) over many billions of years \citep[e.g.,][]{larson+1980, schawinski+2014, peng+2015, lilly+2016}. ETGs, on the other hand, would require their gas reservoir to be exhausted on very short timescales after their morphology has transformed from that of an LTG (e.g., via a major or minor merger). Other processes have also been invoked to explain the transition of galaxies to the red sequence: ram pressure stripping and/or strangulation following the infall of a galaxy into a massive cluster potential \citep{gunn+1972,chung+2009}, tidal disruption and harassment in group interactions \citep{bitsakis+2014}, morphological quenching \citep{martig+2009,martig+2013}, and active galactic nucleus (AGN) feedback \citep{diMatteo+2005,hopkins+2008, fabian2012, kormendy+2013}.

Identifying galaxies in the midst of the transition, as star-formation is being quenched, has not been so straightforward. Traditionally, post-starburst galaxies \citep[also known as E+A or K+A because of the prevalence of A stars in an elliptical galaxy-like spectrum, which contains K stars; ][]{quintero+2004,dressler&gunn1983} have been identified based on the presence of strong Balmer absorption lines, which select for intermediate-age A stars, and the absence of strong star-formation lines, such as H$\alpha$ and [OII]$\lambda 3727$ \citep[e.g.,][]{goto2007}. These criteria are able to select recently quenched galaxies, but they present an incomplete set because they miss objects with line emission from AGN \citep{cales+2013}, objects with substantial emission from post-asymptotic giant branch stars  \citep[post-AGB; ][]{yan+2006}, and shocks \citep{rich+2011,alatalo+2016}, as all of these phenomena can produce non-negligible [OII] and H$\alpha$ emission.

Identifying post-transition galaxies solely photometrically has the potential to revolutionize our understanding of galaxy evolution, removing the necessity of the observationally-expensive follow-up spectroscopy. Notably, post-starburst galaxies exhibit near-uniform optical and near-infrared (NIR) properties \citep{kriek+2010,melnick+2014}. Post-starbursts also exhibit mid-IR colors that place them into a distinct section of color-space \citep{ko+2013,yesuf+2014,alatalo+2017}. While mid-IR colors are effective at low redshift, they become increasingly difficult to observe at high redshift. In contrast, rest-frame UV becomes observable in the optical, making UV observations an excellent tool for identifying and characterizing transitioning galaxies even at high redshift. While rest-frame UV is capable of identifying post-starbursts, directly predicting UV emission in these systems has remained challenging. \citet{melnick+2014} find that stellar population synthesis (SPS) models reproduce the spectral energy distributions (SEDs) of post-starburst galaxies well at optical and near-IR wavelengths, they systematically overpredict the observed UV fluxes of these galaxies. The authors attribute the discrepancy in UV to a combination of inadequate modeling of the synthetic SEDs, and non-uniform distribution of dust leading to an underestimation of the reddening of the intermediate-age populations.

\citet{kaviraj+2007} reconstructed the star formation histories of 38 post-starburst galaxies using UV and optical data from \GALEX\ and SDSS, respectively, and found that the burst of star formation that dominates the post-starburst signature usually takes place within a Gyr, consistent with the presence of A stars. They found that short time scales for the burst (0.01--0.2\,Gyr) and high stellar mass fractions (20--60\%) suggested high SFRs during the burst, resulting in a tight positive correlation with galaxy mass. The SFRs are comparable to those found in luminous and ultra-luminous infrared galaxies (LIRGs and ULIRGs) at low redshift, suggesting that low-redshift massive LIRGs may be the progenitors of massive post-starburst galaxies. The authors were also able to follow the quenching evolution of each modeled post-starburst galaxy, confirming that both the optical and UV colors evolve from blue to red. They also demonstrated that the interrelation between the optical and UV provide further constraints capable of identifying and understanding the evolving galaxy population. The ultraviolet is a wavelength regime that is able to pinpoint galaxies that are currently undergoing that transition. \citet{wild+2014} utilized principle component analysis to differentiate between galaxy types at $ z=0.9-1.2$ based on their SEDs. This analysis was able to directly pinpoint post-starburst galaxies, with the lynchpin wavelength coverage appearing in the UV.

\begin{figure*}[t]
\centering
\includegraphics[width=0.85\linewidth]{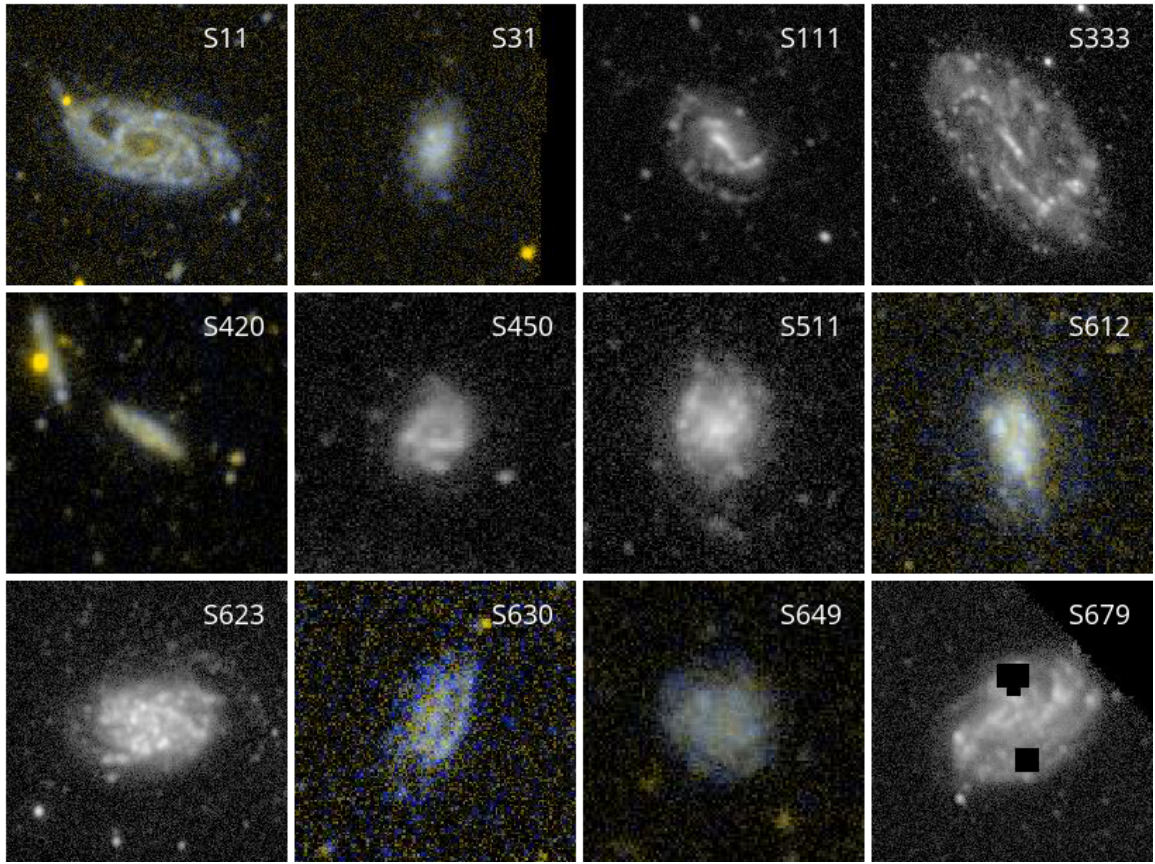}
\caption{\textit{GALEX} UV thumbnails of 12 resolved SPOGs with interesting morphologies. Those in color with yellow text combine FUV data in blue with NUV in yellow. The grayscale images with white text are NUV only. The text at the top right of each image is the SPOGS index number. The bar at the bottom left of each image indicates the scale of one arcminute.}
\label{fig:lowz_lowZ_spogs}
\end{figure*}

Star formation is far from the only source of UV emission present in galaxies. Radiative shock waves induced by supersonic turbulence in gas reservoirs provide an additional source of emission especially relevant for transitioning galaxies. High velocity shocks can ionize the gas and emit highly excited UV emission lines \citep{allen+1998}. Evidence of shocks is frequently seen in merging systems \citep{rich+2011,rich+2014}, galaxies in the outskirts of groups \citep{appleton+2013} and clusters \citep{braglia+2009}, and AGN-dominated galaxies \citep{ogle+2010,villar-martin+2014,lanz+2016}, all processes that may be associated with the quenching of star formation and the transition of blue cloud galaxies into the red sequence. While the emission-line spectra will depend strongly on the physical and ionization structure of the shock, the total luminosity of the shock will effectively only depend on the gas density and shock velocity. Because shocks are characterized by highly-ionized regions of high electron temperature, their spectra can contain several collisionally excited UV lines \citep{allen+2008}, capable of contaminating broadband measurements.

\begin{figure}[t!]
\centering
\includegraphics[width=\linewidth]{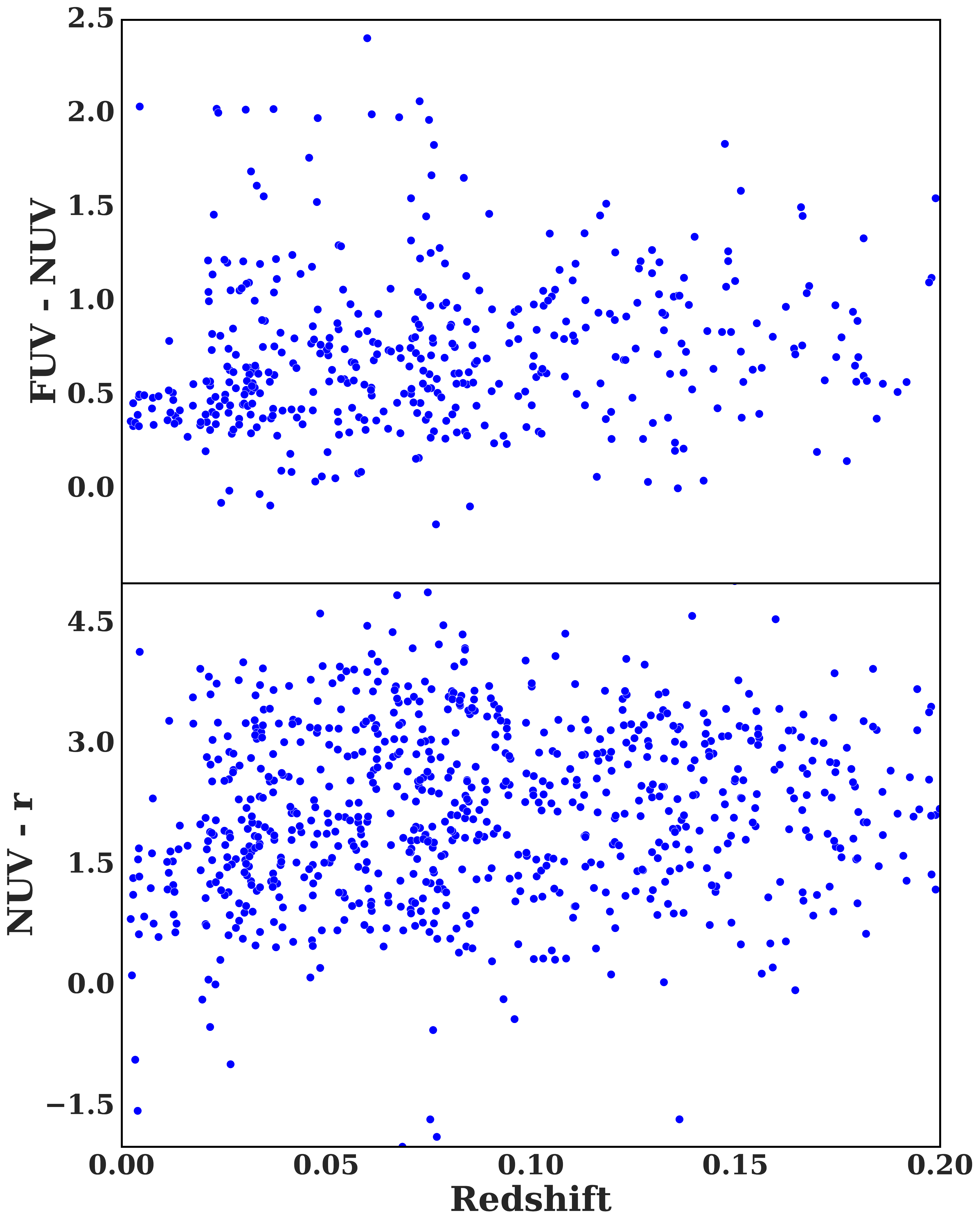}
\caption{Top: k-corrected FUV--NUV color vs. redshift of SPOGs. Bottom: k-corrected NUV--\textit{r} color vs. redshift of SPOGs. In both cases, mean k-corrected color remains relatively constant across redshifts.}
\label{fig:k-corr}
\end{figure}

Traditionally, discriminating between different excitation mechanisms has relied on emission line ratios, usually of optical wavelengths. Most notably, ratios of $\text{[NII]}\lambda 6583/\text{H}\alpha$ vs. $\text{[OIII]} \lambda 5007/\text{H}\beta$, $\text{[SII]}\lambda\lambda 6717,31/\text{H}\alpha$ vs. $\text{[OIII]} \lambda 5007/\text{H}\beta$, and $\text{[OI]}\lambda 6300/\text{H}\alpha$ vs. $\text{[OIII]} \lambda 5007/\text{H}\beta$ have been used to distinguish between star-forming galaxies, composite AGN/HII  galaxies, Seyferts and low-ionization nuclear emission-line regions \citep[LINERs;][]{heckman1986}, and objects excited by shock-wave heating, in line diagnostic diagrams \citep[sometimes referred to as ``BPT/VO87 diagrams"; ][]{baldwin+1981, veilleux&osterbrock1987, kewley+2006}. \citet{allen+1998} showed that UV emission lines also provide a useful way to discern the presence of shocks in galaxies, particularly NIII $\lambda 991$, NIII] $\lambda 1750$, CIII $\lambda 977$, CIV $\lambda 1549$, CIII] $\lambda 1909$, CII] $\lambda 2326$, and HeII $\lambda 1640$.

The Shocked POststarburst Galaxies Survey \citep[SPOGS; ][]{alatalo+2016}\footnote{http://www.spogs.org} seeks to identify galaxies in the midst of transformation overlooked by classical selection criteria, with nebular emission lines excited by shocks rather than star formation. The SPOGS selection criteria are formally defined in \S2 of \cite{alatalo+2016}. Briefly, the catalog selects galaxies from the SDSS DR7 \citep{abazajian+2009} with $z<0.2$, using the Oh-Sarzi-Schawinski-Yi (OSSY) absorption and emission line catalog \citep{oh+2011} to determine line strengths. Following continuum and emission line S/N cuts to ensure robust detections of spectral lines, a criterion is used to select strong Balmer absorption lines for a stellar population with a recent burst of star formation \citep{falkenberg+2009}. Specifically,  we use a threshold of EW(H$\delta$)$>5\angstrom$, consistent with the Balmer post-starburst selection criteria of \citet{goto2007} and \citet{falkenberg+2009}. Next, shock boundaries are defined based on grids of shock models generated from MAPPINGS III \citep{dopita+1995} from the following optical emission line ratios: [NII]/H$\alpha$, [SII]/H$\alpha$, [OI]/H$\alpha$, and [OIII]/H$\beta$ (see \citeauthor{alatalo+2016} \citeyear{alatalo+2016} for details). The remaining galaxies are subclassified based on their line diagnostic ratios as either Seyferts,  LINERS, composites, star-forming, or ambiguous according to the classification lines of \citet{kewley+2006}. In order to limit contamination, we exclude galaxies that are classified as star-forming or composite in all of the three line diagnostic diagrams. The final SPOGS catalog contains the 1067 objects (which we refer to as the SPOGs sample) that meet the aforementioned criteria.

In \S\ref{sec:GALEX}, we explain our methods for obtaining \GALEX UV and SDSS optical photometry for our SPOGs sample, as well as our comparison samples. In \S\ref{sec:analysis}, we present the UV properties of SPOGs. In \S\ref{sec:discussion}, we discuss and provide interpretations for the large scatter in UV photometry and blue UV colors the SPOGs. In \S\ref{sec:summary}, we provide a summary. The cosmological parameters H$_0 = 70$ km\,s$^{-1}$\,Mpc$^{-1}$, $\Omega_M = 0.3$ and $\Omega_{\Lambda} = 0.7$ are assumed throughout \citep{spergel+2007}. \newline

\section{The GALEX SPOG sample}\label{sec:GALEX}
 \GALEX has two UV filters: the far-ultraviolet (FUV; 1350--1750\AA) centered at 1530\AA\ and the near-ultraviolet (NUV; 1750--2750\AA) centered at 2310\AA. We obtained \textit{GALEX} photometry from the Catalog Archive Server Jobs System (CasJobs)\footnote{http://skyserver.sdss.org/casjobs/}, which provides access to the \GALEX\ Data Release 6 (GR6) object catalogs \citep{bianchi+14b}. We used the ``fuv\_mag" and ``nuv\_mag" magnitudes, which correspond to the flux within elliptical ``Kron" apertures \citep[with semimajor axis scaled to 2.5 times the first moment of each source's radial profile, as first suggested by][]{kron+1980}. We also visually inspected intensity maps in both FUV and NUV, to ensure only robust detections (i.e., undeniably present in the image) were included in the analysis. Although most detections were only discernible point sources, \textit{GALEX} did successfully resolve a few objects, 12 of which are shown in Figure~\ref{fig:lowz_lowZ_spogs}. Our selection and inspection resulted in 444 SPOGs detected in both bands, 365 detected in NUV only, and 24 detected only in FUV, for a total of 833 SPOGs with UV detections. The other 234 SPOGs were not observed by \textit{GALEX} in either band. We used the same SDSS Data Release 9 \citep{ahn+2012} photometry ({\em u,\,g,\,r,\,i,\,z} bands) to be consistent with \citet{alatalo+2016}. \newline

\subsection{Comparison Samples}\label{sec:comparison}
We compared SPOGs to samples of Star-forming and Quiescent galaxies selected from \citet{chang+2015}. These samples were defined by their SFR in relation to the star formation main sequence: the Star-forming sample had SFRs within one standard deviation of the star-formation main sequence, Quiescents had SFRs more than five standard deviations below the star-formation main sequence (we refer to them as the SF and Q samples respectively). We also compared to: E+A galaxies selected from SDSS based on the presence of strong Balmer absorption lines and the absence of major emission lines \citep[which we refer to as the E+A sample; ][]{goto2007}; merging galaxies taken from the IRAS Bright Galaxy Sample \citep[which we refer to as the Interacting sample; ][]{sanders+03}; and galaxies with an X-ray-selected AGN \citep[which we refer to as the AGN sample; ][]{lamassa+13}. For each of these samples, SDSS and \GALEX\ photometry were obtained in the same way as with SPOGs, by querying CasJobs.

\subsection{Photometric Corrections}
For all of the comparison samples, SDSS magnitudes are taken from DR9, the same photometry used for SPOGs.  Both \GALEX\ and SDSS magnitudes are corrected for Galactic extinction using the \citet{schlafly+2011} extinction map provided by the NASA/IPAC Extragalactic Database (NED), assuming a \citet{fitzpatrick1999} reddening law, R$_V=A_{V}/E(B-V) = 3.1$, and \textit{GALEX} extinction laws: $A_{NUV}/E(B-V)=8.741$ and $A_{FUV}/E(B-V)=8.376$ from \citet{wyder+2005}. Corrections for intrinsic extinction are done in two ways. For the AGN and Interacting samples, which do not have $E(B-V)$ values from the OSSY catalog \citep{oh+2011}, we apply the analytic formulae of \citet{cho+09}, using the SDSS isophotal axis ratios ($a/b$) of the 25 mag arcsec$^{-2}$ isophote in the \textit{i}-band, and concentration indices ($c=R_{90}/R_{50}$). Specifically, we use Eq. 18 in \cite{cho+09} to calculate the extinction-corrected \textit{r}-band absolute magnitude:
$$
M_{r,0} = -20.77 + \frac{-1+\sqrt[]{1+4\Delta(M_{r,obs}+20.77+4.93\Delta)}}{2\Delta},
$$
where:
$$
\Delta \equiv 1.06 \times 0.223 \times [1.35(c-2.48)^{2} - 1.14]\log_{10}(a/b) \leq 0 .
$$
And we calculate the intrinsic extinction in the \textit{r}-band as $ A_{r,in}=M_{r,0}-M_{r,obs}$.

For the remaining samples, both SDSS and \GALEX\ magnitudes are corrected for intrinsic extinction using the stellar $E(B-V)$ values from the OSSY catalog \citep{oh+2011} and SDSS extinction values from \citet{schlafly+2011}. \citet{oh+2011} calculate their reddening values using \citeauthor{sarzi+2006}'s (\citeyear{sarzi+2006}) \texttt{gandalf} code to match the stellar continuum and nebular emission of each galaxy to various templates. Their models include two reddening components: one which represents dust diffusion in the entire spectrum, and another which only takes into account nebular emission. We used the former for our calculations.

All photometric magnitudes are k-corrected using the ``k-corrections calculator" Python script\footnote{http://kcor.sai.msu.ru/} \citep{chilingarian+2010,chilingarian+2012}.  To check our k-corrections, we plotted k-corrected  NUV--\textit{r} and FUV--NUV colors against redshift and found no apparent dependence on redshift for either color, suggesting the corrections are indeed sufficiently robust (Fig.~\ref{fig:k-corr}).

\begin{figure}[t!]
\centering
\includegraphics[width=\linewidth]{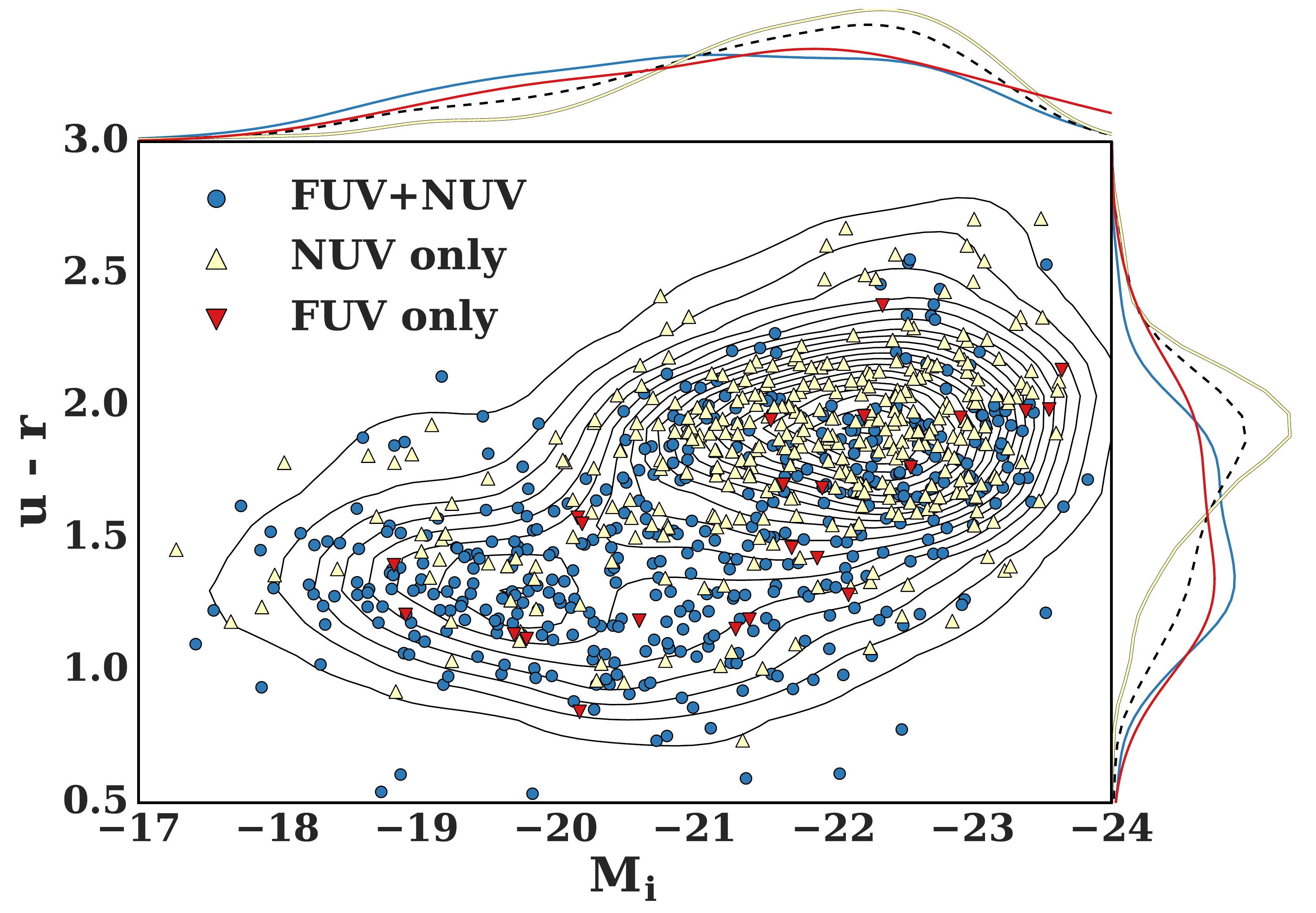}
\caption{SDSS \textit{u--r} color vs. absolute \textit{i}-band magnitude (a proxy for stellar mass) of SPOGs with UV coverage. The contours in the background show the distribution of all 1067 SPOGs. Histograms that collapse the distributions along the {\em u--r} (right) and $M_i$ (top) axes are shown as well. Blue circles show SPOGs with both FUV and NUV coverage, yellow triangles show SPOGs that only have coverage in NUV, and red triangles are SPOGs that only have FUV coverage. Objects detected in both FUV+NUV filters select optically bluer SPOGs, though the effect is not substantial.}
\label{fig:optical_CMD}
\end{figure}

\subsection{Sample Coverage}
\begin{figure}[t!]
\centering
\includegraphics[width=\linewidth]{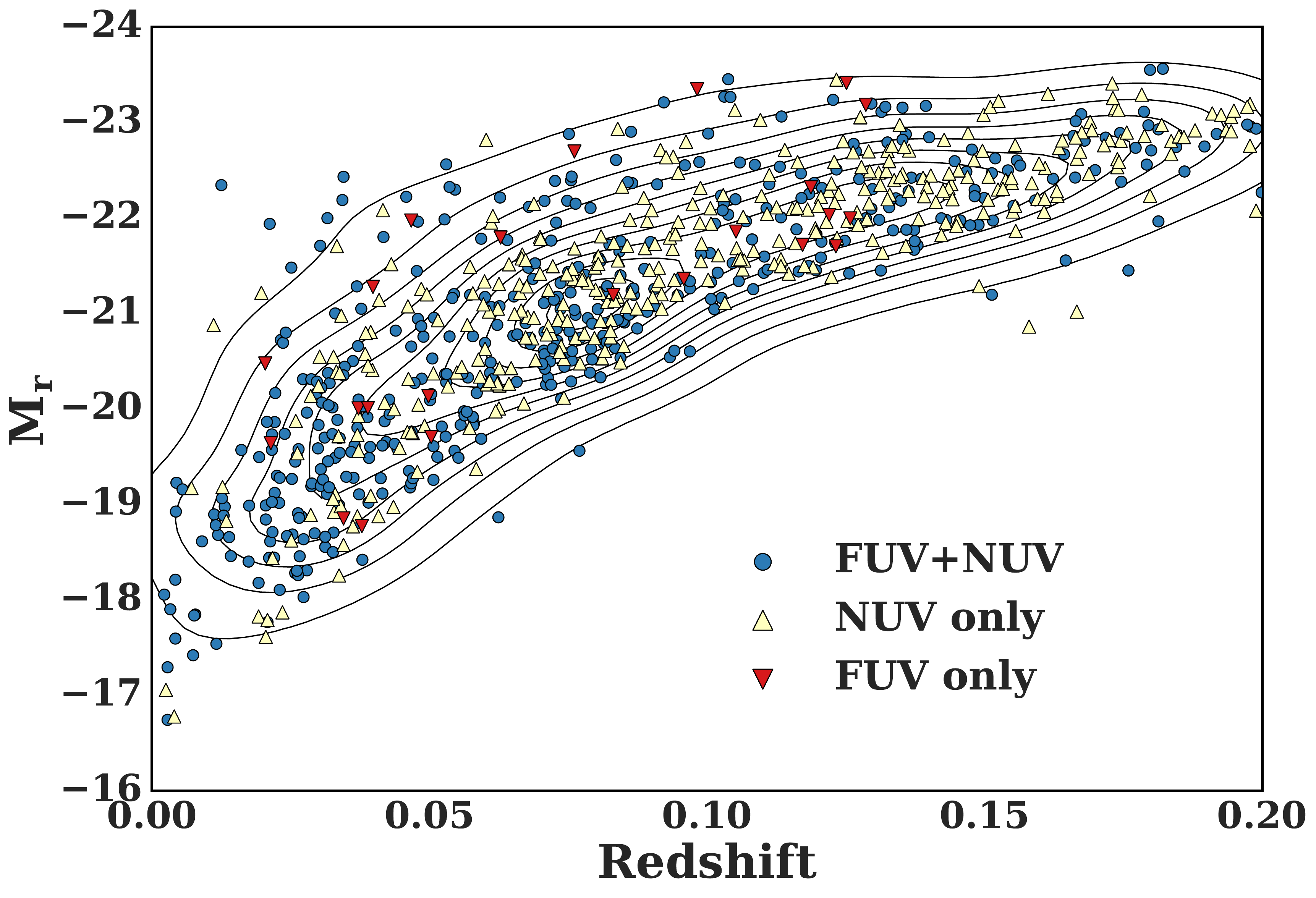}
\caption{SDSS \textit{r}-band absolute magnitude vs. redshift of SPOGs with UV coverage. The contours in the background show the distribution of all 1067 SPOGs. Blue circles show SPOGs with both FUV and NUV coverage, yellow triangles show SPOGs that only have coverage in NUV, and red triangles are SPOGs that only have FUV coverage. SPOGs with both FUV and NUV photometry have a similar parameter coverage in this space to all SPOGs, and but the coverage of UV SPOGs in this parameter space becomes even more complete when SPOGs with only NUV photometry are included.}
\label{fig:mag_redshift}
\end{figure}

\begin{figure}[t!]
\centering
\includegraphics[width=\linewidth]{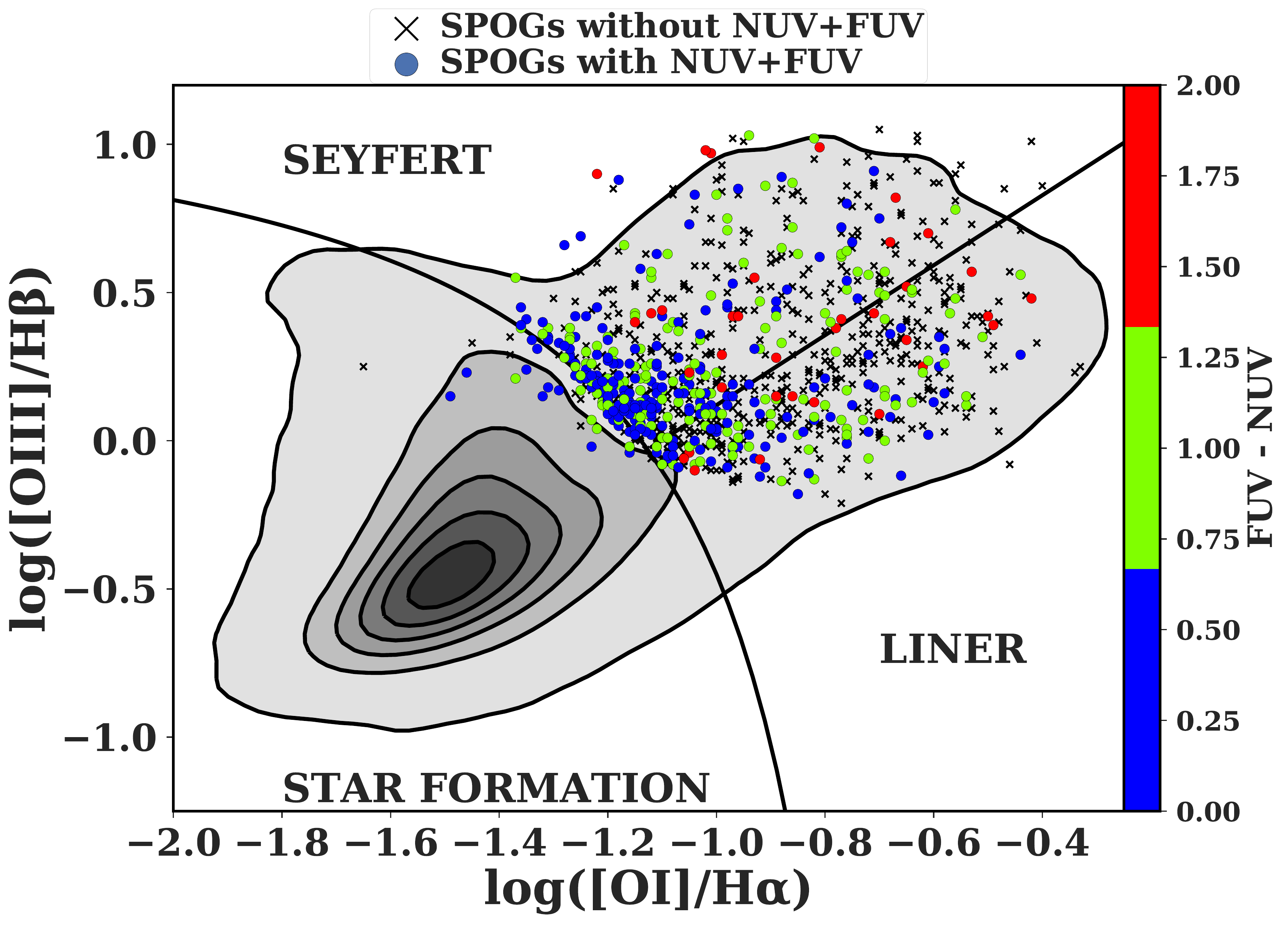}
\caption{Comparison of FUV--NUV colors of SPOGS to their locations on an [OI]/H$\alpha$ emission-line diagram. The gray contours show the emission line galaxy (ELG) sample of \citet{alatalo+2014}, spaced in increments of 10 percentiles of the maximum density. The black x's show SPOGs without both NUV and FUV data, while the colored circular markers show SPOGs with both NUV and FUV coverage, where the color of the marker indicates the FUV--NUV color of the SPOG object. We observe that most of the SPOGs that cluster around the boundary between the star-formation, LINER, and Seyfert regions of the diagram (but mostly fall inside the Seyfert region) generally have bluer UV colors compared to the entire sample.}
\label{fig:FUV-NUV_BPT}
\end{figure}

To determine how representative the set of 833 SPOGs with \textit{GALEX} UV observations are compared to the full sample of 1067, we performed multiple checks by plotting UV SPOGs against the full sample in optical color-magnitude diagrams (Fig.~\ref{fig:optical_CMD}), optical magnitude-redshift space (Fig.~\ref{fig:mag_redshift}), and emission-line ratio diagnostic diagrams (Fig.~\ref{fig:FUV-NUV_BPT}). We find that the UV SPOGs populate a similar parameter space to the full SPOGs sample. In optical color-magnitude space (Fig.~\ref{fig:optical_CMD}), we see that SPOGs with both FUV and NUV photometry are slightly bluer than the full SPOGs sample, which could bias our discussion to galaxies at lower redshift with more signatures of star formation, potentially biasing our conclusions. The sample becomes more  representative when SPOGs with only NUV photometry are included. Based on Anderson Darling (AD) tests \citep{anderson&darling1952}, the distributions of \textit{u--r} colors for NUV-only SPOGs and FUV+NUV SPOGs are distinct ($p<10^{-4}$). Similarly, the distribution of NUV--\textit{r} colors is also distinct between the two samples ($p<10^{-5}$). In order to check for possible biases in the UV properties of the optically bluer sample of FUV+NUV SPOGs, we tested subsamples of FUV+NUV SPOGs that had statistically similar distributions of \textit{u--r} colors as the full SPOGs sample (using Anderson-Darling tests and $p > 0.05$). Using three of these distinct subsamples of 100 FUV+NUV SPOGs, we found no significant biases in UV colors or emission line ratios for any of these subsamples when compared to the full SPOGs sample. We performed a similar test for NUV--\textit{r} colors, where we matched the NUV--\textit{r} distribution of FUV+NUV SPOGS to the full sample of SPOGs with NUV detection and found no significant biases in UV colors or emission line ratios.

In magnitude-redshift space (Fig.~\ref{fig:mag_redshift}), FUV+NUV SPOGs have a similar coverage to the full sample of SPOGs, especially when including SPOGs with only NUV photometry as well. In emission-line ratio space (Fig.~\ref{fig:FUV-NUV_BPT}),
we notice the same trend, that SPOGs with UV photometry occupy a similar space to the full SPOGs sample. These checks indicate that the UV properties of the \textit{GALEX}-detected subsample is representative of the full sample of SPOGs.

\begin{table}
	\centering
	\begin{tabular}{lcc}
		\hline
		Sample & NUV--\textit{r} & FUV--NUV \\
		\hline
		SPOGs & $2.18 \pm 1.15$ & $0.75 \pm 0.55$\\
		Quiescent & $4.54 \pm 1.00$ & $0.74 \pm 0.58$\\
       Star-forming & $2.01 \pm 0.68$ & $0.64 \pm 0.34$\\
        E+A & $3.51 \pm 0.65$ & $1.17 \pm 0.47$\\
        AGN & $2.48 \pm 1.24$ & $0.67 \pm 0.40$\\
        Interacting & $2.03 \pm 1.66
$ & $0.86 \pm 0.47
$\\
        \hline
	\end{tabular}
    \caption{NUV--\textit{r} and FUV--NUV colors of all of our samples. For each sample the mean color is given as well as the standard deviation. }
	\label{tab:colors}
\end{table}

\section{Analysis}\label{sec:analysis}
\subsection{Color-Magnitude Space}\label{sec:c-m space}
\begin{figure*}
\centering
\subfigure{\includegraphics[trim={2.3cm 6cm 3.5cm 0.cm},clip, width=\linewidth]{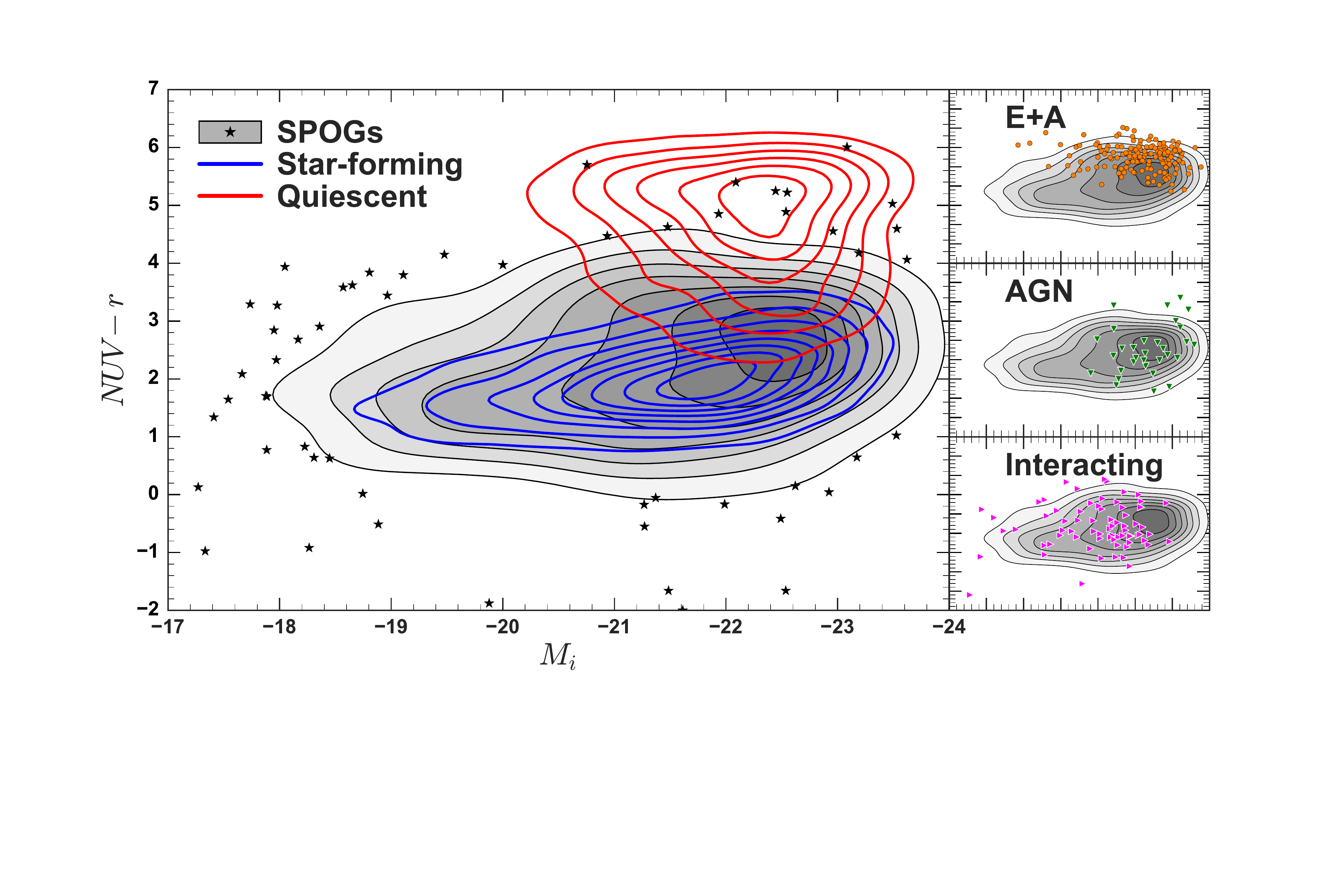}}
\subfigure{\includegraphics[trim={2.3cm 6cm 3.5cm 0.cm},clip, width=\linewidth]{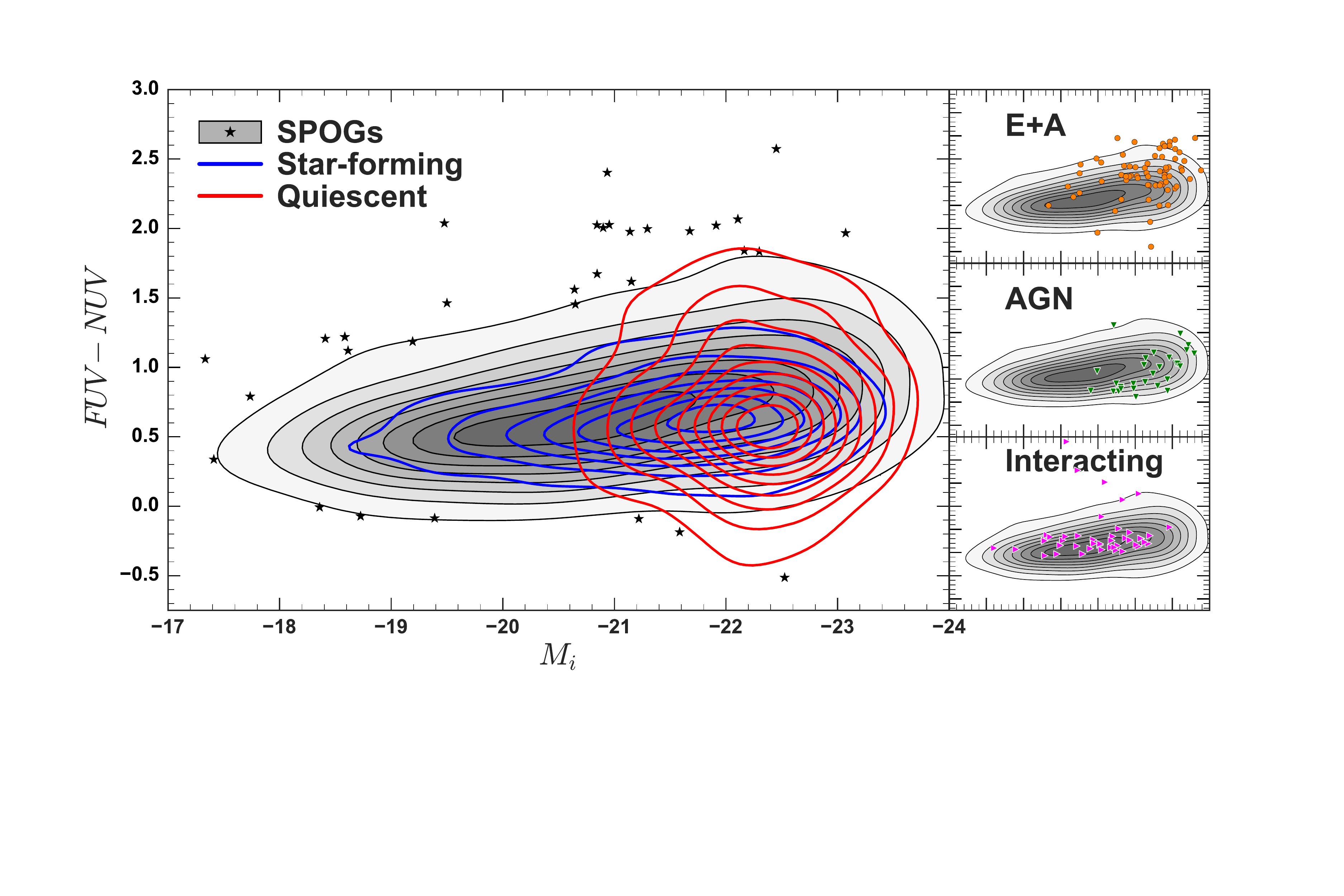}}
\caption{Top: k-corrected NUV--\textit{r} color vs. SDSS \textit{i}-band absolute magnitude ($M_i$) of SPOGS (gray contours) and all comparison samples. Bottom: k-corrected FUV--NUV color vs. SDSS \textit{i}-band absolute magnitude ($M_i$) of SPOGS (gray contours) and all comparison samples. In both plots, the black stars are SPOGs that fall outside the outer contour. While most of the comparison samples fall within the color-magnitude space occupied by SPOGs, no single sample fully covers the SPOGs parameter space, suggesting that SPOGs may be composed of a combination of these samples. All contours are in increments of 10 percentiles of the maximum density}.
\label{fig:cmd}
\end{figure*}

In Fig.~\ref{fig:cmd}, we show the distribution of SPOGs in NUV--\textit{r} vs. $M_i$ and FUV--NUV versus $M_i$ space compared to galaxies that are Star-forming, Quiescent, E+A, AGN hosts, and Interacting. In NUV--\textit{r}, SPOGs occupy a region similar to the Star-forming sample (and distinct from Quiescents). The E+A, AGN, and Interacting samples occupy distinct regions of NUV--\textit{r} space, and all share some overlap with SPOGs, suggesting that SPOGs are a heterogeneous group comprised of galaxies with different mechanisms for UV emission. In FUV--NUV (bottom panel of Fig. \ref{fig:cmd}), we find a similar trend, with SPOGs displaying a larger range in FUV--NUV colors compared to the Star-forming sample. Once again the E+A, AGN, and Interacting samples all share a similar color distribution as SPOGs, but seem to occupy distinct subregions. These plots suggest that UV emission in SPOGs may be an amalgam of the mechanisms that the comparison samples represent in different proportions. Future multiwavelength and high-resolution morphological analyses will clarify which individual SPOGs exhibit similar UV emission mechanisms to each comparison group.

\begin{figure}[t!]
\centering
\includegraphics[width=\linewidth]{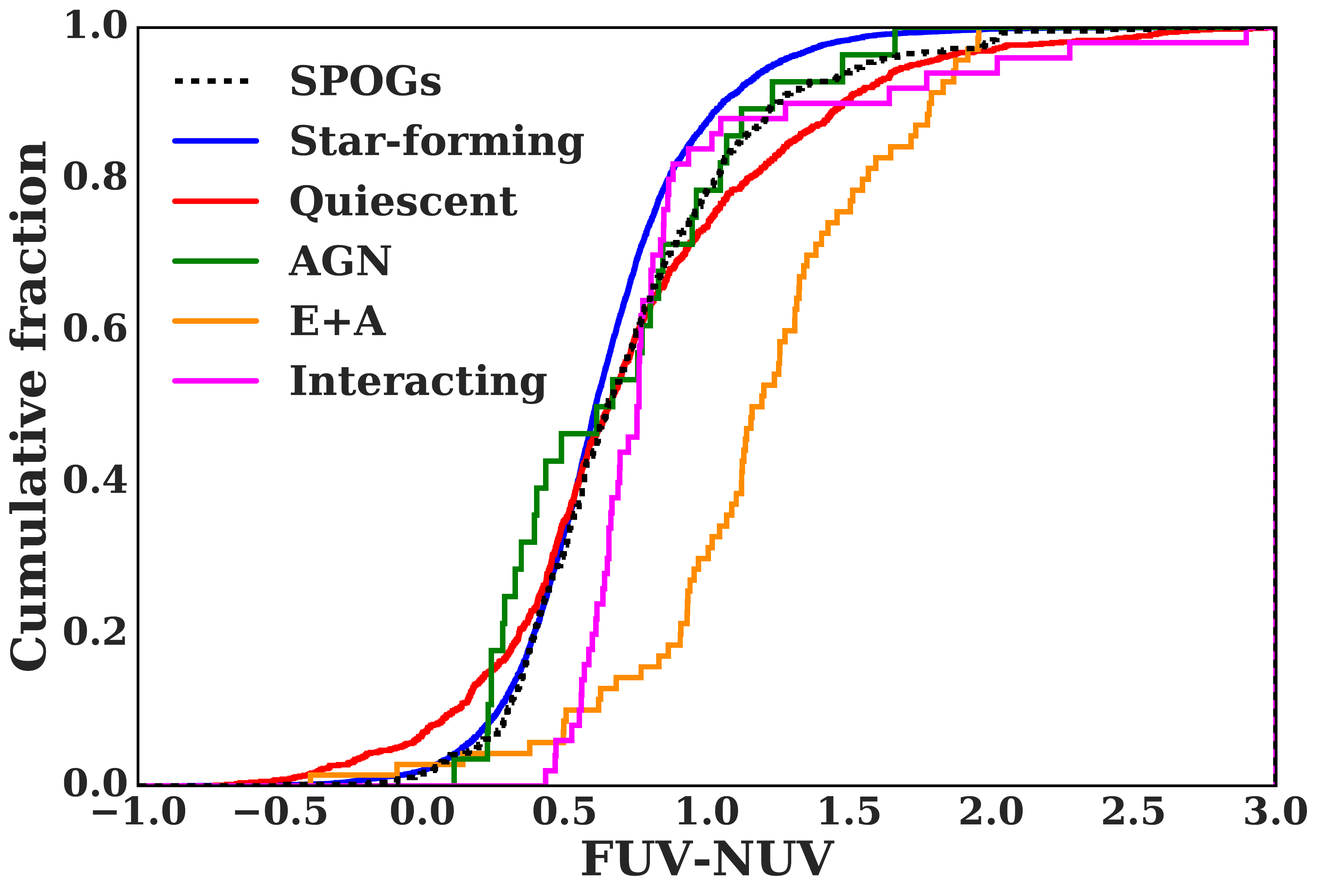}
\caption{Cumulative distributions of \textit{GALEX} FUV--NUV color for SPOGs and each of our comparison samples. With Anderson-Darling tests, we show that SPOGs are a distinct distribution from all other comparison samples, except AGN (see \S ~\ref{sec:c-m space})}
\label{fig:cdf}
\end{figure}

Fig.~\ref{fig:cdf} shows the cumulative distribution of FUV--NUV colors for SPOGs compared to each of the comparison samples. We used AD tests to compare the shapes of these 1D UV color distributions. For samples in the full range of masses, all samples, except the AGN sample, have statistically distinct color distributions from SPOGs with greater than 99\% significance. The AGN sample is distinct from SPOGs at the 78\% significance level. When restricting the comparison to a specific range of masses ($-23 < M_i < -21$), all samples, except the Interacting sample, have statistically distinct color distributions from SPOGs with greater than 99\% significance. The Interacting sample is distinct from SPOGs at the 85\% significance level. This test shows that the differences in UV colors are not attributable to differences in stellar mass distribution.

\subsection{Mixture Models}\label{sec:mixtures}
In order to further test which of the comparison samples contribute to the FUV--NUV colors of SPOGs and in what fractions, we used the method of Gaussian mixture models (GMMs). GMM models some underlying distribution whose true probability density function is unknown as the sum multiple Gaussian distributions \citep{ivezic+2014}. Here we first modeled each of our comparison samples as a mixture of Gaussian components, then we modeled our SPOGs distribution as a mixture of our comparison samples. Specifically, we only compared SPOGs to SFs, Qs, and AGN because to zeroth-order, these constitute a base set of galaxy types. Interacting galaxies constitute a heterogeneous set composed of different kinds of galaxies, so we did not include them in our mixture models in order to keep our results about the composition of SPOGs simple and clear.

\begin{figure}[t!]
\centering
\includegraphics[width=\linewidth]{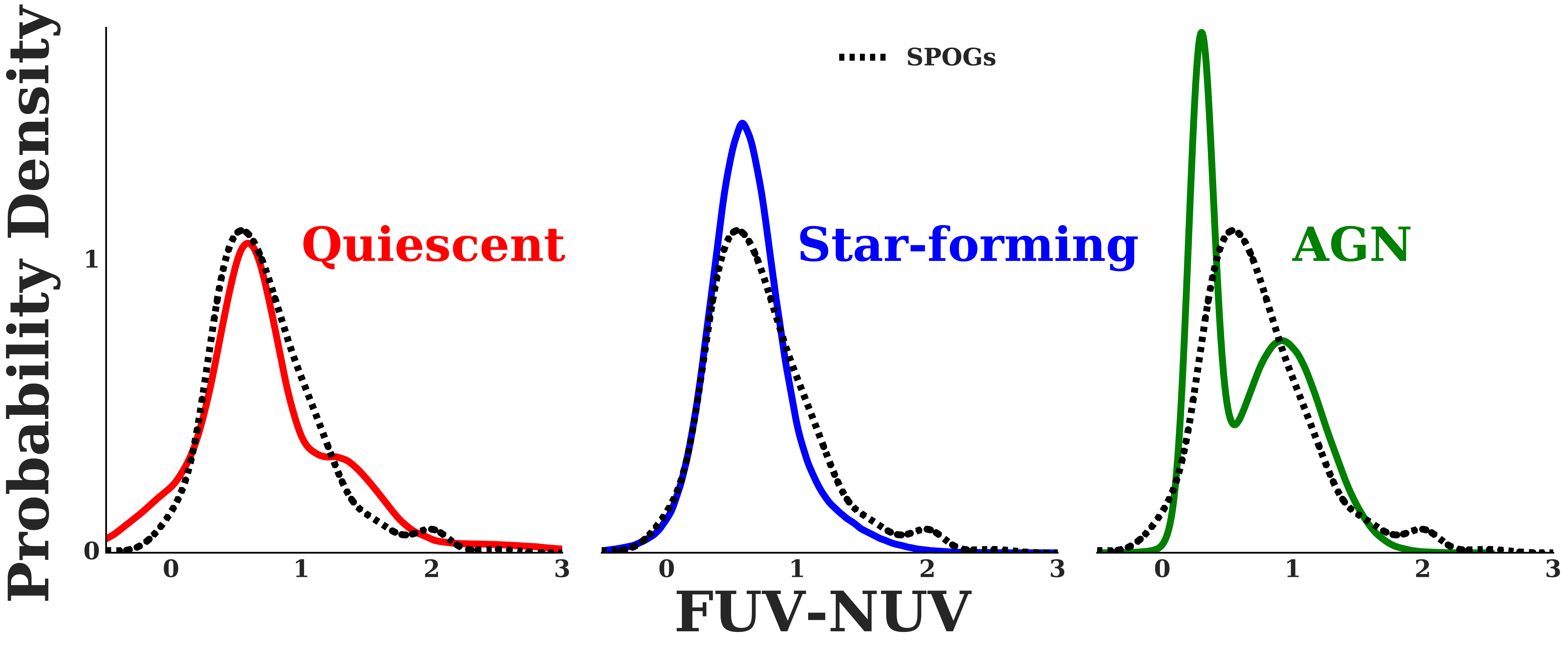}
\caption{The three components of our mixture models used to replicate the SPOGs distribution: Quiescent, Star-forming, and AGN. The distribution of Quiescents is best reproduced by a mixture of four Gaussian distributions, while the Star-forming and AGN are each sufficiently reproduced by a two-Gaussian model (see \S ~\ref{sec:c-m space}). The distribution of SPOGs (reproduced using a three-Gaussian model) is shown in black dashed lines.}
\label{fig:shapes}
\end{figure}

We approximated the distribution of FUV--NUV colors of the SFs, Qs, and AGN comparison samples as mixtures of Gaussian distributions. In order to fit GMMs to each of our samples, we used an expectation-maximization algorithm \citep{dempster+1977}, which iteratively adjusts the parameters of each Gaussian component until a maximum likelihood is reached for a given sample.  This was done for GMMs with different number of components, and the Bayes Information Criterion (BIC) was used to select the optimal number of components for each comparison sample. The SF and AGN samples were modeled as two-component Gaussian mixtures, while the sample of Qs required a mixture of four Gaussians. We also modeled the SPOGs distribution as a three-component Gaussian mixture, and E+A as a single component Gaussian, which we used to compare to our mixtures of components. In Fig.~\ref{fig:shapes}, we show the smoothed shapes of the FUV--NUV distributions of SFs, Qs, AGN, and SPOGs. We note that no single comparison distribution has a shape identical to that of the SPOGs distribution. As a result, we attempt to mix different fractions of SFs, Qs, and AGN to match the SPOGs distribution.

We created mixture models with every two-component combination of these three base distributions (SFs+Qs, SFs+AGN, and Qs+AGN). For each mixture model, we varied only the weights of the two components, constrained by the fact that the sum of the weights had to be unity. A grid search method was used to search the parameter space (weights had to be between 0 and 1) for the likelihood of different mixture weights. At each weight, a likelihood was calculated as the sum of the probability density function of the GMM over all data points we were trying to fit. The grid size was of 1000 values.

\begin{figure*}[t]
\subfigure{\includegraphics[height=4.1cm]{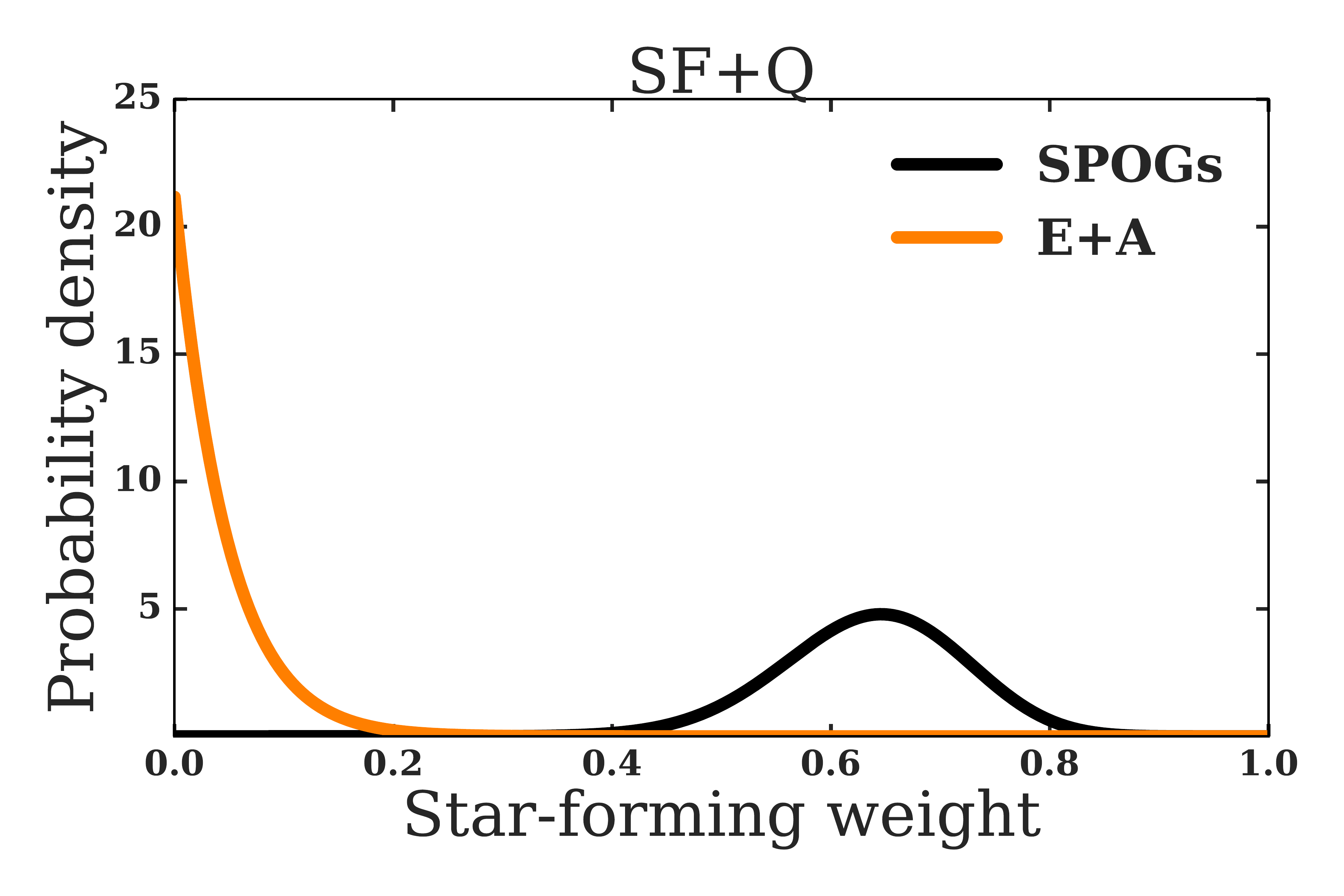}}
\subfigure{\includegraphics[height=4.1cm]{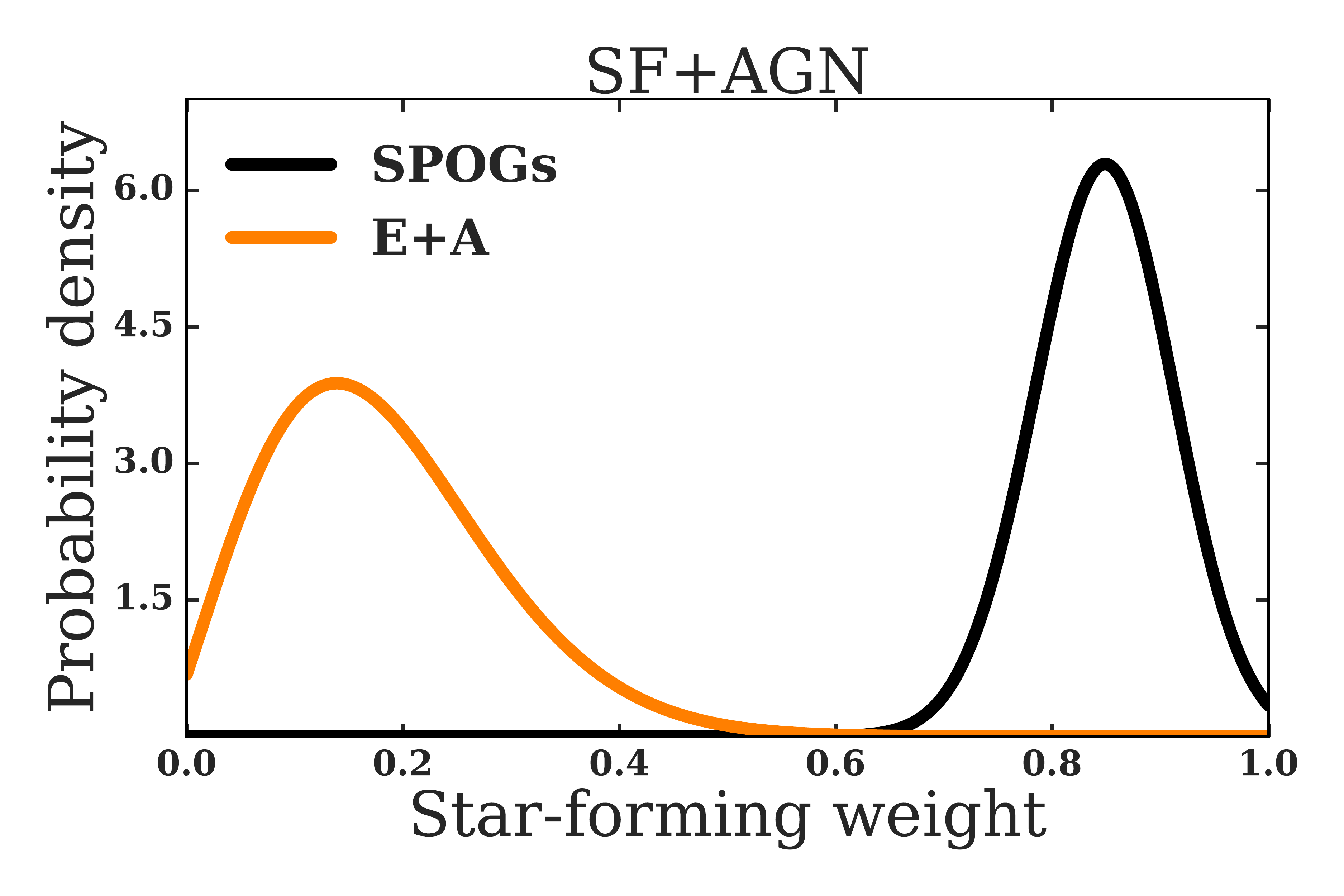}}
\subfigure{\includegraphics[height=4.1cm]{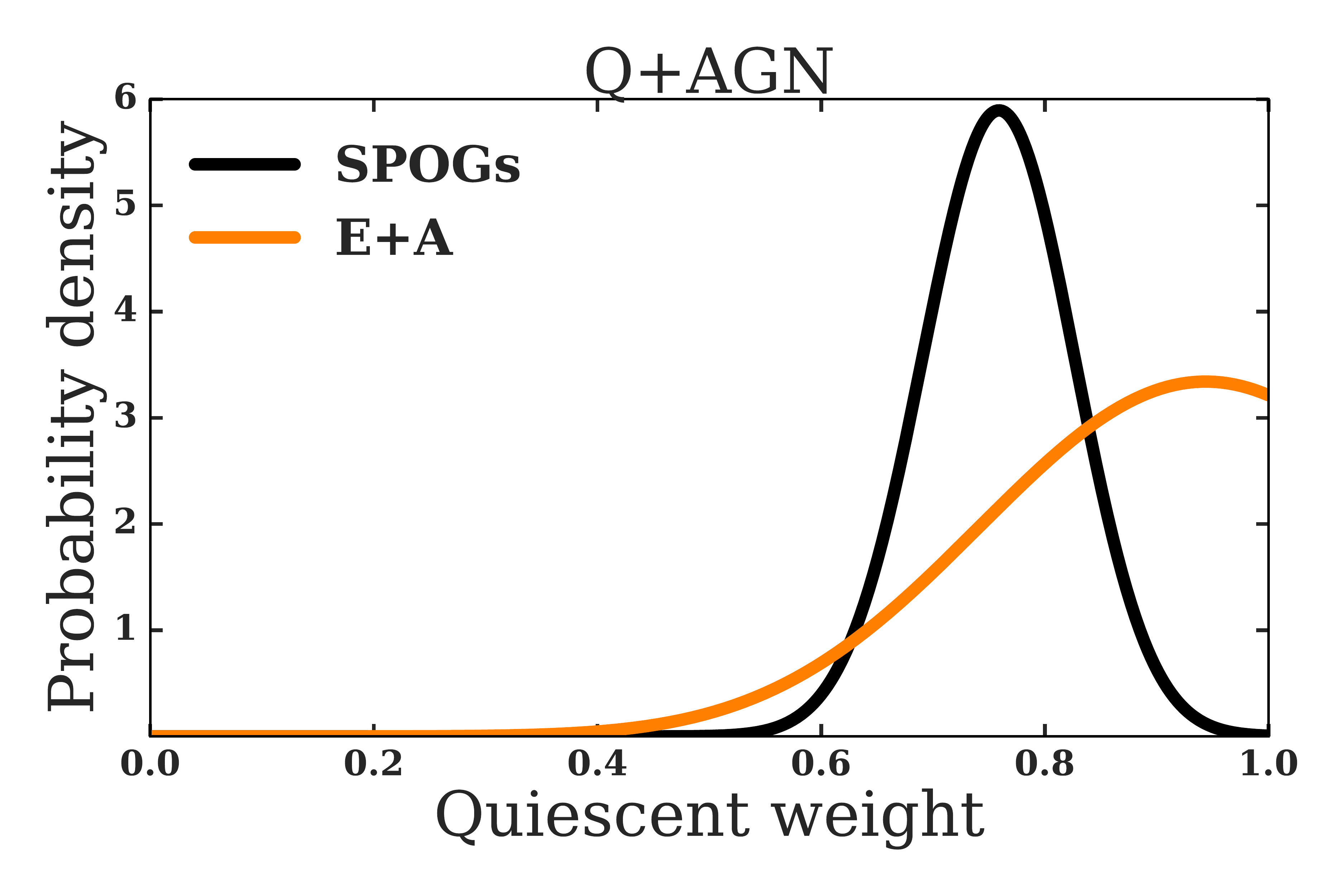}}
\caption{The results of performing the minimization procedure described in \S~\ref{sec:mixtures} on the two-component mixture models for both SPOGs (in black) and the E+A sample (in orange). The mixture models differ significantly between SPOGs and E+A galaxies, with E+A galaxies requiring a much more significant contribution of ETGs to explain their UV colors than SPOGs.}
\label{fig:mixture_model}
\end{figure*}

We perform this grid search for both SPOGs and E+A galaxies and show the results in Fig.~\ref{fig:mixture_model}. These plots show the potential contribution of each component to SPOGs and E+As.  We observe that the SPOGs distribution is best reproduced with a large fraction of Star-forming galaxies ($>60\%$), and a small fraction of AGN galaxies ($<30\%$). E+A galaxies require a large fraction of Quiescent galaxies ($>90\%$) and a smaller fraction of Star-forming ($<20\%$) galaxies. The differences in the contributions of the Quiescent and Star-forming populations in the models for SPOGs and E+A galaxies supports the idea that E+A galaxies are at a later stage of galaxy evolution (i.e. more quiesced star formation) than SPOGs.

\section{Discussion}\label{sec:discussion}
\subsection{SPOG FUV--NUV Colors}
As Figures \ref{fig:cmd}b and \ref{fig:cdf} show, SPOGs exhibit average FUV--NUV colors that are consistent with Star-forming objects, and redder than average E+A galaxies and Quiescents. The E+A population has the most consistently red UV colors, compared to other populations, including Quiescents. In fact, there is a pronounced blue FUV--NUV wing amongst Quiescents. This blue wing in the distribution of Quiescents is likely due to the UV upturn phenomenon, in which horizontal branch stars are thought to provide sufficient FUV emission in massive quiescent galaxies to create an observable bump \citep{yi+97}. Given that this phenomenon occurs only in the most massive quiescent galaxies and occurs amongst $\sim$\,10\,Gyr old stars, it is unsurprising that it does not appear in other distributions (including the E+As), as it takes an extended period of time for these stars to become a significant source of NUV and FUV emission. Our selection criteria also require a UV detection in at least one band to be a part of the Quiescent comparison sample, so the presence of a non-negligible number of UV-upturn Quiescent galaxies is unsurprising.

The population that is the closest match to SPOGs is the Star-forming population, thus we surmise that the most likely origin of the UV emission in SPOGs is their stellar population.  $31\%$ of SPOGs have blue UV colors (FUV--NUV$<0.5$), compared with $9\%$ of E+A galaxies, $46\%$ of AGN, $6\%$ of Interacting, $34\%$ of Star-forming, and $35\%$ of Quiescents. The larger fraction of blue UV colors further provides support for the idea that SPOGs are in an earlier stage in the transition from blue cloud to red sequence compared to E+A galaxies.

\subsection{Heterogeneity in the UV Colors of SPOGs}
The large scatters in the NUV--\textit{r} and FUV--NUV colors (Fig.~\ref{fig:cmd}), as well as in the $M_i$ absolute magnitudes\footnote{ This scatter is likely in part due to Malmquist bias, and likely is related to the large redshift range that is being probed for SPOGs \citep{alatalo+2016}.} \citep[a proxy for galaxy mass; ][]{bell+2003}, suggest that SPOGs are a heterogeneous group, possibly composed of different classes of objects. \citet{melnick+2014} reached a similar conclusion about E+A galaxies, suggesting that variability in dust obscuration also contributes to the heterogeneity. From color-magnitude diagrams, it is plausible to suggest that the full sample of SPOGs contains objects that belong to each of the comparison samples: Star-forming, E+A, AGN, and Interacting, with a small contribution of Quiescents. This heterogeneity may even be unsurprising, given the broad criterion the SPOG survey was based on: shock-like line ratios in the ionized gas combined with Balmer absorption from intermediate-aged stars. There are many types of objects that may be able to fit this description, from low metallicity dwarf galaxies whose ionized gas line ratios fall outside of the star-forming region defined by \citet{kewley+2006}, to rejuvenated early-type galaxies \citep{alatalo+2016b}, to AGN host galaxies \citep{cales+2011}. The UV colors also suggest that the SPOG criterion identifies physical mechanisms (shocks + intermediate-aged stars), regardless of the galaxy in which said physical mechanism is hosted.

The UV-color GMMs support the idea that E+A galaxies are more advanced in their transition than SPOGS, since their UV colors are most similar to Quiescents. Thus, the UV colors of E+A galaxies compared to SPOGS support the optical \citep{melnick+2014,alatalo+2016} and infrared \citep{alatalo+2014,alatalo+2017} observations that SPOGs are ``younger'' than E+A galaxies. Therefore, combined with optical and infrared evidence, our UV studies suggest that SPOGs as a population are in an earlier stage of their transition from blue star-formers to red quiescents. An analysis of the full SEDs of SPOGs will provide a more detailed differentiation between ``types'' of SPOGs, and the relative contributions of each comparison sample, but this is beyond the scope of this paper (Bitsakis et al. in prep).

\section{Summary}\label{sec:summary}
In this paper we investigated the UV properties of SPOGs compared to Quiescent, Star-forming, E+A galaxies, AGN-dominated galaxies, and Interacting galaxies. We found that:

\begin{itemize}
  \item SPOGs show a larger scatter in their FUV--NUV and NUV--\textit{r} colors compared to most of the other samples, although Quiescent, Star-forming, E+A galaxies, AGN, and Interacting galaxies all occupy overlapping UV color parameter space similar to SPOGs. This result suggests that SPOGs are a heterogeneous group, possibly including several of these classes of objects.

  \item SPOGs exhibit FUV--NUV colors that are consistent with star-forming galaxies, and are much bluer than E+A galaxies, suggesting that they are at an earlier stage of their transition from blue to red.

  \item We used Gaussian mixture models to attempt to measure the relative contribution of Quiescent, Star-forming, and AGN hosts to the UV colors of SPOGs, and find that they require a  $>$60\% contribution of Star-forming, and a small fraction of AGN population ($<$30\%) .

  \item We ran Gaussian mixture models on E+A galaxies and found that their UV colors required a much smaller contribution of Star-forming, supporting both the optical and infrared picture that SPOGs are earlier in their transition than E+A galaxies.
\end{itemize}

Analyzing the full SEDs of SPOGs will allow us to obtain a better understanding of stellar populations, masses, star formation histories, and radiation sources. Obtaining UV spectral data of these objects will also be necessary to fully understand the UV emission. Ultimately, understanding the UV emission in SPOGs will allow us to better place them within the context of galaxy evolution.

\acknowledgments FA thanks Mark Seibert and Michael A. Strauss for their guidance and comments to this work; Andy Skemer for introducing some of the statistical techniques that permitted the analysis; and the anonymous referee for their informative and insightful report and suggestions.

The following software and programming languages made this research possible: \texttt{PYTHON} (v2.7); \texttt{ASTROPY} \citep[v2.0;][]{astropy1, astropy2}, a community-developed core \texttt{PYTHON} package for Astronomy; \texttt{Scikit-learn} \citep[v0.19;][]{scikit-learn}; \texttt{SEABORN} (v0.8.1;); and Tool for OPerations on Catalogues And Tables \citep[TOPCAT, v4.3;][]{topcat}.

Support for this project was made possible by the collaboration between Princeton University and the Observatories of the Carnegie Institution. Support for KA is provided by NASA through Hubble Fellowship grant \hbox{\#HST-HF2-51352.001} awarded by the Space Telescope Science Institute, which is operated by the Association of Universities for Research in Astronomy, Inc., for NASA, under contract NAS5-26555. Support for AMM is provided by NASA through Hubble Fellowship grant \hbox{\#HST-HF2-51377} awarded by the Space Telescope Science Institute, which is operated by the Association of Universities for Research in Astronomy, Inc., for NASA, under contract NAS5-26555. TB would like to acknowledge support from the CONACyT Research Fellowships program. KN acknowledges support from the grant associated with Spitzer proposal 11086. J.~F-B. acknowledges support from grant AYA2016-77237-C3-1-P from the Spanish Ministry of Economy and Competitiveness (MINECO).
This research has made use of the NASA/IPAC Extragalactic Database (NED) which is operated by the Jet Propulsion Laboratory, California Institute of Technology, under contract with the National Aeronautics and Space Administration.

\GALEX\ (Galaxy Evolution Explorer) is a NASA Small Explorer, launched in 2003 April. We gratefully acknowledge NASA's support for construction, operation, and science analysis for the \GALEX\ mission, developed in cooperation with the Centre National d'Etudes Spatiales of France and the Korean Ministry of Science and Technology. Funding for SDSS-III has been provided by the Alfred P. Sloan Foundation, the Participating Institutions, the National Science Foundation, and the U.S. Department of Energy Office of Science. The SDSS-III web site is http://www.sdss3.org/.

{\it Facility:} \facility{GALEX}, \facility{Sloan}
\bibliographystyle{aasjournal}
\bibliography{master}

\begin{thebibliography}{}
\expandafter\ifx\csname natexlab\endcsname\relax\def\natexlab#1{#1}\fi

\bibitem[{{Abazajian} {et~al.}(2009){Abazajian}, {Adelman-McCarthy},
  {Ag{\"u}eros}, {Allam}, {Allende Prieto}, {An}, {Anderson}, {Anderson},
  {Annis}, {Bahcall}, \& et~al.}]{abazajian+2009}
{Abazajian}, K.~N., {Adelman-McCarthy}, J.~K., {Ag{\"u}eros}, M.~A., {et~al.}
  2009, \apjs, 182, 543

\bibitem[{{Ahn} {et~al.}(2012){Ahn}, {Alexandroff}, {Allende Prieto},
  {Anderson}, {Anderton}, {Andrews}, {Aubourg}, {Bailey}, {Balbinot}, {Barnes},
  \& et~al.}]{ahn+2012}
{Ahn}, C.~P., {Alexandroff}, R., {Allende Prieto}, C., {et~al.} 2012, \apjs,
  203, 21

\bibitem[{{Alatalo} {et~al.}(2014){Alatalo}, {Cales}, {Appleton}, {Kewley},
  {Lacy}, {Lisenfeld}, {Nyland}, \& {Rich}}]{alatalo+2014}
{Alatalo}, K., {Cales}, S.~L., {Appleton}, P.~N., {et~al.} 2014, \apjl, 794,
  L13

\bibitem[{{Alatalo} {et~al.}(2016{\natexlab{a}}){Alatalo}, {Lisenfeld}, {Lanz},
  {Appleton}, {Ardila}, {Cales}, {Kewley}, {Lacy}, {Medling}, {Nyland}, {Rich},
  \& {Urry}}]{alatalo+2016b}
{Alatalo}, K., {Lisenfeld}, U., {Lanz}, L., {et~al.} 2016{\natexlab{a}}, \apj,
  827, 106

\bibitem[{{Alatalo} {et~al.}(2016{\natexlab{b}}){Alatalo}, {Cales}, {Rich},
  {Appleton}, {Kewley}, {Lacy}, {Lanz}, {Medling}, \& {Nyland}}]{alatalo+2016}
{Alatalo}, K., {Cales}, S.~L., {Rich}, J.~A., {et~al.} 2016{\natexlab{b}},
  \apjs, 224, 38

\bibitem[{{Alatalo} {et~al.}(2017){Alatalo}, {Bitsakis}, {Lanz}, {Lacy},
  {Brown}, {French}, {Ciesla}, {Appleton}, {Beaton}, {Cales}, {Crossett},
  {Falc{\'o}n-Barroso}, {Kelson}, {Kewley}, {Kriek}, {Medling}, {Mulchaey},
  {Nyland}, {Rich}, \& {Urry}}]{alatalo+2017}
{Alatalo}, K., {Bitsakis}, T., {Lanz}, L., {et~al.} 2017, \apj, 843, 9

\bibitem[{{Allen} {et~al.}(1998){Allen}, {Dopita}, \& {Tsvetanov}}]{allen+1998}
{Allen}, M.~G., {Dopita}, M.~A., \& {Tsvetanov}, Z.~I. 1998, \apj, 493, 571

\bibitem[{{Allen} {et~al.}(2008){Allen}, {Groves}, {Dopita}, {Sutherland}, \&
  {Kewley}}]{allen+2008}
{Allen}, M.~G., {Groves}, B.~A., {Dopita}, M.~A., {Sutherland}, R.~S., \&
  {Kewley}, L.~J. 2008, \apjs, 178, 20

\bibitem[{Anderson \& Darling(1952)}]{anderson&darling1952}
Anderson, T.~W., \& Darling, D.~A. 1952, Ann. Math. Statist., 193

\bibitem[{{Appleton} {et~al.}(2013){Appleton}, {Guillard}, {Boulanger},
  {Cluver}, {Ogle}, {Falgarone}, {Pineau des For{\^e}ts}, {O'Sullivan}, {Duc},
  {Gallagher}, {Gao}, {Jarrett}, {Konstantopoulos}, {Lisenfeld}, {Lord}, {Lu},
  {Peterson}, {Struck}, {Sturm}, {Tuffs}, {Valchanov}, {van der Werf}, \&
  {Xu}}]{appleton+2013}
{Appleton}, P.~N., {Guillard}, P., {Boulanger}, F., {et~al.} 2013, \apj, 777,
  66

\bibitem[{{Astropy Collaboration} {et~al.}(2013){Astropy Collaboration},
  {Robitaille}, {Tollerud}, {Greenfield}, {Droettboom}, {Bray}, {Aldcroft},
  {Davis}, {Ginsburg}, {Price-Whelan}, {Kerzendorf}, {Conley}, {Crighton},
  {Barbary}, {Muna}, {Ferguson}, {Grollier}, {Parikh}, {Nair}, {Unther},
  {Deil}, {Woillez}, {Conseil}, {Kramer}, {Turner}, {Singer}, {Fox}, {Weaver},
  {Zabalza}, {Edwards}, {Azalee Bostroem}, {Burke}, {Casey}, {Crawford},
  {Dencheva}, {Ely}, {Jenness}, {Labrie}, {Lim}, {Pierfederici}, {Pontzen},
  {Ptak}, {Refsdal}, {Servillat}, \& {Streicher}}]{astropy1}
{Astropy Collaboration}, {Robitaille}, T.~P., {Tollerud}, E.~J., {et~al.} 2013,
  \aap, 558, A33

\bibitem[{{Baade}(1958)}]{baade1958}
{Baade}, W. 1958, Ricerche Astronomiche, 5, 3

\bibitem[{{Baldry} {et~al.}(2004){Baldry}, {Glazebrook}, {Brinkmann},
  {Ivezi{\'c}}, {Lupton}, {Nichol}, \& {Szalay}}]{baldry+04}
{Baldry}, I.~K., {Glazebrook}, K., {Brinkmann}, J., {et~al.} 2004, \apj, 600,
  681

\bibitem[{{Baldwin} {et~al.}(1981){Baldwin}, {Phillips}, \&
  {Terlevich}}]{baldwin+1981}
{Baldwin}, J.~A., {Phillips}, M.~M., \& {Terlevich}, R. 1981, \pasp, 93, 5

\bibitem[{{Bell} {et~al.}(2003){Bell}, {McIntosh}, {Katz}, \&
  {Weinberg}}]{bell+2003}
{Bell}, E.~F., {McIntosh}, D.~H., {Katz}, N., \& {Weinberg}, M.~D. 2003, \apjs,
  149, 289

\bibitem[{{Bell} {et~al.}(2012){Bell}, {van der Wel}, {Papovich}, {Kocevski},
  {Lotz}, {McIntosh}, {Kartaltepe}, {Faber}, {Ferguson}, {Koekemoer}, {Grogin},
  {Wuyts}, {Cheung}, {Conselice}, {Dekel}, {Dunlop}, {Giavalisco},
  {Herrington}, {Koo}, {McGrath}, {de Mello}, {Rix}, {Robaina}, \&
  {Williams}}]{bell+2012}
{Bell}, E.~F., {van der Wel}, A., {Papovich}, C., {et~al.} 2012, \apj, 753, 167

\bibitem[{{Bianchi} {et~al.}(2014){Bianchi}, {Conti}, \& {Shiao}}]{bianchi+14b}
{Bianchi}, L., {Conti}, A., \& {Shiao}, B. 2014, VizieR Online Data Catalog,
  2335

\bibitem[{{Bitsakis} {et~al.}(2014){Bitsakis}, {Charmandaris}, {Appleton},
  {D{\'{\i}}az-Santos}, {Le Floc'h}, {da Cunha}, {Alatalo}, \&
  {Cluver}}]{bitsakis+2014}
{Bitsakis}, T., {Charmandaris}, V., {Appleton}, P.~N., {et~al.} 2014, \aap,
  565, A25

\bibitem[{{Braglia} {et~al.}(2009){Braglia}, {Pierini}, {Biviano}, \&
  {B{\"o}hringer}}]{braglia+2009}
{Braglia}, F.~G., {Pierini}, D., {Biviano}, A., \& {B{\"o}hringer}, H. 2009,
  \aap, 500, 947

\bibitem[{{Cales} {et~al.}(2011){Cales}, {Brotherton}, {Shang}, {Bennert},
  {Canalizo}, {Stoll}, {Ganguly}, {Vanden Berk}, {Paul}, \&
  {Diamond-Stanic}}]{cales+2011}
{Cales}, S.~L., {Brotherton}, M.~S., {Shang}, Z., {et~al.} 2011, \apj, 741, 106

\bibitem[{{Cales} {et~al.}(2013){Cales}, {Brotherton}, {Shang}, {Runnoe},
  {DiPompeo}, {Bennert}, {Canalizo}, {Hiner}, {Stoll}, {Ganguly}, \&
  {Diamond-Stanic}}]{cales+2013}
---. 2013, \apj, 762, 90

\bibitem[{{Chang} {et~al.}(2015){Chang}, {van der Wel}, {da Cunha}, \&
  {Rix}}]{chang+2015}
{Chang}, Y.-Y., {van der Wel}, A., {da Cunha}, E., \& {Rix}, H.-W. 2015, \apjs,
  219, 8

\bibitem[{{Chilingarian} {et~al.}(2010){Chilingarian}, {Melchior}, \&
  {Zolotukhin}}]{chilingarian+2010}
{Chilingarian}, I.~V., {Melchior}, A.-L., \& {Zolotukhin}, I.~Y. 2010, \mnras,
  405, 1409

\bibitem[{{Chilingarian} \& {Zolotukhin}(2012)}]{chilingarian+2012}
{Chilingarian}, I.~V., \& {Zolotukhin}, I.~Y. 2012, \mnras, 419, 1727

\bibitem[{{Cho} \& {Park}(2009)}]{cho+09}
{Cho}, J., \& {Park}, C. 2009, \apj, 693, 1045

\bibitem[{{Chung} {et~al.}(2009){Chung}, {van Gorkom}, {Kenney}, {Crowl}, \&
  {Vollmer}}]{chung+2009}
{Chung}, A., {van Gorkom}, J.~H., {Kenney}, J.~D.~P., {Crowl}, H., \&
  {Vollmer}, B. 2009, \aj, 138, 1741

\bibitem[{Dempster {et~al.}(1977)Dempster, Laird, \& Rubin}]{dempster+1977}
Dempster, A.~P., Laird, N.~M., \& Rubin, D.~B. 1977, J. Roy. Statist. Soc. Ser.
  B, 39, 1, with discussion

\bibitem[{{Di Matteo} {et~al.}(2005){Di Matteo}, {Springel}, \&
  {Hernquist}}]{diMatteo+2005}
{Di Matteo}, T., {Springel}, V., \& {Hernquist}, L. 2005, \nat, 433, 604

\bibitem[{{Dopita} \& {Sutherland}(1995)}]{dopita+1995}
{Dopita}, M.~A., \& {Sutherland}, R.~S. 1995, \apj, 455, 468

\bibitem[{{Dressler} \& {Gunn}(1983)}]{dressler&gunn1983}
{Dressler}, A., \& {Gunn}, J.~E. 1983, \apj, 270, 7

\bibitem[{{Faber} {et~al.}(2007){Faber}, {Willmer}, {Wolf}, {Koo}, {Weiner},
  {Newman}, {Im}, {Coil}, {Conroy}, {Cooper}, {Davis}, {Finkbeiner}, {Gerke},
  {Gebhardt}, {Groth}, {Guhathakurta}, {Harker}, {Kaiser}, {Kassin},
  {Kleinheinrich}, {Konidaris}, {Kron}, {Lin}, {Luppino}, {Madgwick},
  {Meisenheimer}, {Noeske}, {Phillips}, {Sarajedini}, {Schiavon}, {Simard},
  {Szalay}, {Vogt}, \& {Yan}}]{faber+2007}
{Faber}, S.~M., {Willmer}, C.~N.~A., {Wolf}, C., {et~al.} 2007, \apj, 665, 265

\bibitem[{{Fabian}(2012)}]{fabian2012}
{Fabian}, A.~C. 2012, \araa, 50, 455

\bibitem[{{Falkenberg} {et~al.}(2009){Falkenberg}, {Kotulla}, \&
  {Fritze}}]{falkenberg+2009}
{Falkenberg}, M.~A., {Kotulla}, R., \& {Fritze}, U. 2009, \mnras, 397, 1940

\bibitem[{{Fitzpatrick}(1999)}]{fitzpatrick1999}
{Fitzpatrick}, E.~L. 1999, \pasp, 111, 63

\bibitem[{{Goto}(2007)}]{goto2007}
{Goto}, T. 2007, \mnras, 377, 1222

\bibitem[{{Gunn} \& {Gott}(1972)}]{gunn+1972}
{Gunn}, J.~E., \& {Gott}, III, J.~R. 1972, \apj, 176, 1

\bibitem[{{Harker} {et~al.}(2006){Harker}, {Schiavon}, {Weiner}, \&
  {Faber}}]{harker+2006}
{Harker}, J.~J., {Schiavon}, R.~P., {Weiner}, B.~J., \& {Faber}, S.~M. 2006,
  \apjl, 647, L103

\bibitem[{{Heckman}(1986)}]{heckman1986}
{Heckman}, T.~M. 1986, \pasp, 98, 159

\bibitem[{{Hopkins} {et~al.}(2008){Hopkins}, {Cox}, {Kere{\v s}}, \&
  {Hernquist}}]{hopkins+2008}
{Hopkins}, P.~F., {Cox}, T.~J., {Kere{\v s}}, D., \& {Hernquist}, L. 2008,
  \apjs, 175, 390

\bibitem[{{Hubble}(1926)}]{hubble1926}
{Hubble}, E.~P. 1926, \apj, 64, doi:10.1086/143018

\bibitem[{Ivezic {et~al.}(2014)Ivezic, Connolly, VanderPlas, \&
  Gray}]{ivezic+2014}
Ivezic, Z., Connolly, A.~J., VanderPlas, J.~T., \& Gray, A. 2014, Statistics,
  Data Mining, and Machine Learning in Astronomy: A Practical Python Guide for
  the Analysis of Survey Data (Princeton, NJ, USA: Princeton University Press)

\bibitem[{{Kannappan} {et~al.}(2009){Kannappan}, {Guie}, \&
  {Baker}}]{kannappan+2009}
{Kannappan}, S.~J., {Guie}, J.~M., \& {Baker}, A.~J. 2009, \aj, 138, 579

\bibitem[{{Kaviraj} {et~al.}(2007){Kaviraj}, {Kirkby}, {Silk}, \&
  {Sarzi}}]{kaviraj+2007}
{Kaviraj}, S., {Kirkby}, L.~A., {Silk}, J., \& {Sarzi}, M. 2007, \mnras, 382,
  960

\bibitem[{{Kewley} {et~al.}(2006){Kewley}, {Groves}, {Kauffmann}, \&
  {Heckman}}]{kewley+2006}
{Kewley}, L.~J., {Groves}, B., {Kauffmann}, G., \& {Heckman}, T. 2006, \mnras,
  372, 961

\bibitem[{{Ko} {et~al.}(2013){Ko}, {Hwang}, {Lee}, \& {Sohn}}]{ko+2013}
{Ko}, J., {Hwang}, H.~S., {Lee}, J.~C., \& {Sohn}, Y.-J. 2013, \apj, 767, 90

\bibitem[{{Kormendy} \& {Ho}(2013)}]{kormendy+2013}
{Kormendy}, J., \& {Ho}, L.~C. 2013, \araa, 51, 511

\bibitem[{{Kriek} {et~al.}(2010){Kriek}, {Labb{\'e}}, {Conroy}, {Whitaker},
  {van Dokkum}, {Brammer}, {Franx}, {Illingworth}, {Marchesini}, {Muzzin},
  {Quadri}, \& {Rudnick}}]{kriek+2010}
{Kriek}, M., {Labb{\'e}}, I., {Conroy}, C., {et~al.} 2010, \apjl, 722, L64

\bibitem[{{Kron}(1980)}]{kron+1980}
{Kron}, R.~G. 1980, \apjs, 43, 305

\bibitem[{{LaMassa} {et~al.}(2013){LaMassa}, {Urry}, {Cappelluti}, {Civano},
  {Ranalli}, {Glikman}, {Treister}, {Richards}, {Ballantyne}, {Stern},
  {Comastri}, {Cardamone}, {Schawinski}, {B{\"o}hringer}, {Chon}, {Murray},
  {Green}, \& {Nandra}}]{lamassa+13}
{LaMassa}, S.~M., {Urry}, C.~M., {Cappelluti}, N., {et~al.} 2013, \mnras, 436,
  3581

\bibitem[{{Lanz} {et~al.}(2016){Lanz}, {Ogle}, {Alatalo}, \&
  {Appleton}}]{lanz+2016}
{Lanz}, L., {Ogle}, P.~M., {Alatalo}, K., \& {Appleton}, P.~N. 2016, \apj, 826,
  29

\bibitem[{{Larson} {et~al.}(1980){Larson}, {Tinsley}, \&
  {Caldwell}}]{larson+1980}
{Larson}, R.~B., {Tinsley}, B.~M., \& {Caldwell}, C.~N. 1980, \apj, 237, 692

\bibitem[{{Lilly} \& {Carollo}(2016)}]{lilly+2016}
{Lilly}, S.~J., \& {Carollo}, C.~M. 2016, \apj, 833, 1

\bibitem[{{Lintott} {et~al.}(2008){Lintott}, {Schawinski}, {Slosar}, {Land},
  {Bamford}, {Thomas}, {Raddick}, {Nichol}, {Szalay}, {Andreescu}, {Murray}, \&
  {Vandenberg}}]{lintott+2008}
{Lintott}, C.~J., {Schawinski}, K., {Slosar}, A., {et~al.} 2008, \mnras, 389,
  1179

\bibitem[{{Martig} {et~al.}(2009){Martig}, {Bournaud}, {Teyssier}, \&
  {Dekel}}]{martig+2009}
{Martig}, M., {Bournaud}, F., {Teyssier}, R., \& {Dekel}, A. 2009, \apj, 707,
  250

\bibitem[{{Martig} {et~al.}(2013){Martig}, {Crocker}, {Bournaud}, {Emsellem},
  {Gabor}, {Alatalo}, {Blitz}, {Bois}, {Bureau}, {Cappellari}, {Davies},
  {Davis}, {Dekel}, {de Zeeuw}, {Duc}, {Falc{\'o}n-Barroso}, {Khochfar},
  {Krajnovi{\'c}}, {Kuntschner}, {Morganti}, {McDermid}, {Naab}, {Oosterloo},
  {Sarzi}, {Scott}, {Serra}, {Griffin}, {Teyssier}, {Weijmans}, \&
  {Young}}]{martig+2013}
{Martig}, M., {Crocker}, A.~F., {Bournaud}, F., {et~al.} 2013, \mnras, 432,
  1914

\bibitem[{{Martin} {et~al.}(2005){Martin}, {Fanson}, {Schiminovich},
  {Morrissey}, {Friedman}, {Barlow}, {Conrow}, {Grange}, {Jelinsky},
  {Milliard}, {Siegmund}, {Bianchi}, {Byun}, {Donas}, {Forster}, {Heckman},
  {Lee}, {Madore}, {Malina}, {Neff}, {Rich}, {Small}, {Surber}, {Szalay},
  {Welsh}, \& {Wyder}}]{martin+2005}
{Martin}, D.~C., {Fanson}, J., {Schiminovich}, D., {et~al.} 2005, \apjl, 619,
  L1

\bibitem[{{Melnick} \& {De Propris}(2014)}]{melnick+2014}
{Melnick}, J., \& {De Propris}, R. 2014, \apss, 354, 65

\bibitem[{{Noeske} {et~al.}(2007){Noeske}, {Weiner}, {Faber}, {Papovich},
  {Koo}, {Somerville}, {Bundy}, {Conselice}, {Newman}, {Schiminovich}, {Le
  Floc'h}, {Coil}, {Rieke}, {Lotz}, {Primack}, {Barmby}, {Cooper}, {Davis},
  {Ellis}, {Fazio}, {Guhathakurta}, {Huang}, {Kassin}, {Martin}, {Phillips},
  {Rich}, {Small}, {Willmer}, \& {Wilson}}]{noeske+2007}
{Noeske}, K.~G., {Weiner}, B.~J., {Faber}, S.~M., {et~al.} 2007, \apjl, 660,
  L43

\bibitem[{{Ogle} {et~al.}(2010){Ogle}, {Boulanger}, {Guillard}, {Evans},
  {Antonucci}, {Appleton}, {Nesvadba}, \& {Leipski}}]{ogle+2010}
{Ogle}, P., {Boulanger}, F., {Guillard}, P., {et~al.} 2010, \apj, 724, 1193

\bibitem[{{Oh} {et~al.}(2011){Oh}, {Sarzi}, {Schawinski}, \& {Yi}}]{oh+2011}
{Oh}, K., {Sarzi}, M., {Schawinski}, K., \& {Yi}, S.~K. 2011, \apjs, 195, 13

\bibitem[{Pedregosa {et~al.}(2011)Pedregosa, Varoquaux, Gramfort, Michel,
  Thirion, Grisel, Blondel, Prettenhofer, Weiss, Dubourg, Vanderplas, Passos,
  Cournapeau, Brucher, Perrot, \& Duchesnay}]{scikit-learn}
Pedregosa, F., Varoquaux, G., Gramfort, A., {et~al.} 2011, Journal of Machine
  Learning Research, 12, 2825

\bibitem[{{Peng} {et~al.}(2015){Peng}, {Maiolino}, \& {Cochrane}}]{peng+2015}
{Peng}, Y., {Maiolino}, R., \& {Cochrane}, R. 2015, \nat, 521, 192

\bibitem[{{Quintero} {et~al.}(2004){Quintero}, {Hogg}, {Blanton}, {Schlegel},
  {Eisenstein}, {Gunn}, {Brinkmann}, {Fukugita}, {Glazebrook}, \&
  {Goto}}]{quintero+2004}
{Quintero}, A.~D., {Hogg}, D.~W., {Blanton}, M.~R., {et~al.} 2004, \apj, 602,
  190

\bibitem[{{Rich} {et~al.}(2011){Rich}, {Kewley}, \& {Dopita}}]{rich+2011}
{Rich}, J.~A., {Kewley}, L.~J., \& {Dopita}, M.~A. 2011, \apj, 734, 87

\bibitem[{{Rich} {et~al.}(2014){Rich}, {Kewley}, \& {Dopita}}]{rich+2014}
---. 2014, \apjl, 781, L12

\bibitem[{{Sanders} {et~al.}(2003){Sanders}, {Mazzarella}, {Kim}, {Surace}, \&
  {Soifer}}]{sanders+03}
{Sanders}, D.~B., {Mazzarella}, J.~M., {Kim}, D.-C., {Surace}, J.~A., \&
  {Soifer}, B.~T. 2003, \aj, 126, 1607

\bibitem[{{Sarzi} {et~al.}(2006){Sarzi}, {Falc{\'o}n-Barroso}, {Davies},
  {Bacon}, {Bureau}, {Cappellari}, {de Zeeuw}, {Emsellem}, {Fathi},
  {Krajnovi{\'c}}, {Kuntschner}, {McDermid}, \& {Peletier}}]{sarzi+2006}
{Sarzi}, M., {Falc{\'o}n-Barroso}, J., {Davies}, R.~L., {et~al.} 2006, \mnras,
  366, 1151

\bibitem[{{Schawinski} {et~al.}(2014){Schawinski}, {Urry}, {Simmons},
  {Fortson}, {Kaviraj}, {Keel}, {Lintott}, {Masters}, {Nichol}, {Sarzi},
  {Skibba}, {Treister}, {Willett}, {Wong}, \& {Yi}}]{schawinski+2014}
{Schawinski}, K., {Urry}, C.~M., {Simmons}, B.~D., {et~al.} 2014, \mnras, 440,
  889

\bibitem[{{Schlafly} \& {Finkbeiner}(2011)}]{schlafly+2011}
{Schlafly}, E.~F., \& {Finkbeiner}, D.~P. 2011, \apj, 737, 103

\bibitem[{{Spergel} {et~al.}(2007){Spergel}, {Bean}, {Dor{\'e}}, {Nolta},
  {Bennett}, {Dunkley}, {Hinshaw}, {Jarosik}, {Komatsu}, {Page}, {Peiris},
  {Verde}, {Halpern}, {Hill}, {Kogut}, {Limon}, {Meyer}, {Odegard}, {Tucker},
  {Weiland}, {Wollack}, \& {Wright}}]{spergel+2007}
{Spergel}, D.~N., {Bean}, R., {Dor{\'e}}, O., {et~al.} 2007, \apjs, 170, 377

\bibitem[{{Strateva} {et~al.}(2001){Strateva}, {Ivezi{\'c}}, {Knapp},
  {Narayanan}, {Strauss}, {Gunn}, {Lupton}, {Schlegel}, {Bahcall}, {Brinkmann},
  {Brunner}, {Budav{\'a}ri}, {Csabai}, {Castander}, {Doi}, {Fukugita}, {Gy{\H
  o}ry}, {Hamabe}, {Hennessy}, {Ichikawa}, {Kunszt}, {Lamb}, {McKay},
  {Okamura}, {Racusin}, {Sekiguchi}, {Schneider}, {Shimasaku}, \&
  {York}}]{strateva+01}
{Strateva}, I., {Ivezi{\'c}}, {\v Z}., {Knapp}, G.~R., {et~al.} 2001, \aj, 122,
  1861

\bibitem[{{Taylor}(2005)}]{topcat}
{Taylor}, M.~B. 2005, in Astronomical Society of the Pacific Conference Series,
  Vol. 347, Astronomical Data Analysis Software and Systems XIV, ed.
  P.~{Shopbell}, M.~{Britton}, \& R.~{Ebert}, 29

\bibitem[{{The Astropy Collaboration} {et~al.}(2018){The Astropy
  Collaboration}, {Price-Whelan}, {Sip{\H o}cz}, {G{\"u}nther}, {Lim},
  {Crawford}, {Conseil}, {Shupe}, {Craig}, {Dencheva}, {Ginsburg},
  {VanderPlas}, {Bradley}, {P{\'e}rez-Su{\'a}rez}, {de Val-Borro}, {Aldcroft},
  {Cruz}, {Robitaille}, {Tollerud}, {Ardelean}, {Babej}, {Bachetti}, {Bakanov},
  {Bamford}, {Barentsen}, {Barmby}, {Baumbach}, {Berry}, {Biscani}, {Boquien},
  {Bostroem}, {Bouma}, {Brammer}, {Bray}, {Breytenbach}, {Buddelmeijer},
  {Burke}, {Calderone}, {Cano Rodr{\'{\i}}guez}, {Cara}, {Cardoso},
  {Cheedella}, {Copin}, {Crichton}, {D{\'A}vella}, {Deil}, {Depagne},
  {Dietrich}, {Donath}, {Droettboom}, {Earl}, {Erben}, {Fabbro}, {Ferreira},
  {Finethy}, {Fox}, {Garrison}, {Gibbons}, {Goldstein}, {Gommers}, {Greco},
  {Greenfield}, {Groener}, {Grollier}, {Hagen}, {Hirst}, {Homeier}, {Horton},
  {Hosseinzadeh}, {Hu}, {Hunkeler}, {Ivezi{\'c}}, {Jain}, {Jenness}, {Kanarek},
  {Kendrew}, {Kern}, {Kerzendorf}, {Khvalko}, {King}, {Kirkby}, {Kulkarni},
  {Kumar}, {Lee}, {Lenz}, {Littlefair}, {Ma}, {Macleod}, {Mastropietro},
  {McCully}, {Montagnac}, {Morris}, {Mueller}, {Mumford}, {Muna}, {Murphy},
  {Nelson}, {Nguyen}, {Ninan}, {N{\"o}the}, {Ogaz}, {Oh}, {Parejko}, {Parley},
  {Pascual}, {Patil}, {Patil}, {Plunkett}, {Prochaska}, {Rastogi}, {Reddy
  Janga}, {Sabater}, {Sakurikar}, {Seifert}, {Sherbert}, {Sherwood-Taylor},
  {Shih}, {Sick}, {Silbiger}, {Singanamalla}, {Singer}, {Sladen}, {Sooley},
  {Sornarajah}, {Streicher}, {Teuben}, {Thomas}, {Tremblay}, {Turner},
  {Terr{\'o}n}, {van Kerkwijk}, {de la Vega}, {Watkins}, {Weaver}, {Whitmore},
  {Woillez}, \& {Zabalza}}]{astropy2}
{The Astropy Collaboration}, {Price-Whelan}, A.~M., {Sip{\H o}cz}, B.~M.,
  {et~al.} 2018, ArXiv e-prints, arXiv:1801.02634

\bibitem[{{Tinsley}(1978)}]{tinsley1978}
{Tinsley}, B.~M. 1978, \apj, 222, 14

\bibitem[{{Veilleux} \& {Osterbrock}(1987)}]{veilleux&osterbrock1987}
{Veilleux}, S., \& {Osterbrock}, D.~E. 1987, \apjs, 63, 295

\bibitem[{{Villar Mart{\'{\i}}n} {et~al.}(2014){Villar Mart{\'{\i}}n},
  {Emonts}, {Humphrey}, {Cabrera Lavers}, \& {Binette}}]{villar-martin+2014}
{Villar Mart{\'{\i}}n}, M., {Emonts}, B., {Humphrey}, A., {Cabrera Lavers}, A.,
  \& {Binette}, L. 2014, \mnras, 440, 3202

\bibitem[{{Wild} {et~al.}(2014){Wild}, {Rosales-Ortega}, {Falc{\'o}n-Barroso},
  {Garc{\'{\i}}a-Benito}, {Gallazzi}, {Gonz{\'a}lez Delgado}, {Bekerait{\'e}},
  {Pasquali}, {Johansson}, {Garc{\'{\i}}a Lorenzo}, {van de Ven}, {Pawlik},
  {Per{\'e}z}, {Monreal-Ibero}, {Lyubenova}, {Cid Fernandes},
  {M{\'e}ndez-Abreu}, {Barrera-Ballesteros}, {Kehrig}, {Iglesias-P{\'a}ramo},
  {Bomans}, {M{\'a}rquez}, {Johnson}, {Kennicutt}, {Husemann}, {Mast},
  {S{\'a}nchez}, {Walcher}, {Alves}, {Aguerri}, {Alonso Herrero},
  {Bland-Hawthorn}, {Catal{\'a}n-Torrecilla}, {Florido}, {Gomes}, {Jahnke},
  {L{\'o}pez-S{\'a}nchez}, {de Lorenzo-C{\'a}ceres}, {Marino},
  {M{\'a}rmol-Queralt{\'o}}, {Olden}, {del Olmo}, {Papaderos}, {Quirrenbach},
  {V{\'{\i}}lchez}, \& {Ziegler}}]{wild+2014}
{Wild}, V., {Rosales-Ortega}, F., {Falc{\'o}n-Barroso}, J., {et~al.} 2014,
  \aap, 567, A132

\bibitem[{{Wyder} {et~al.}(2005){Wyder}, {Treyer}, {Milliard}, {Schiminovich},
  {Arnouts}, {Budav{\'a}ri}, {Barlow}, {Bianchi}, {Byun}, {Donas}, {Forster},
  {Friedman}, {Heckman}, {Jelinsky}, {Lee}, {Madore}, {Malina}, {Martin},
  {Morrissey}, {Neff}, {Rich}, {Siegmund}, {Small}, {Szalay}, \&
  {Welsh}}]{wyder+2005}
{Wyder}, T.~K., {Treyer}, M.~A., {Milliard}, B., {et~al.} 2005, \apjl, 619, L15

\bibitem[{{Yan} {et~al.}(2006){Yan}, {Newman}, {Faber}, {Konidaris}, {Koo}, \&
  {Davis}}]{yan+2006}
{Yan}, R., {Newman}, J.~A., {Faber}, S.~M., {et~al.} 2006, \apj, 648, 281

\bibitem[{{Yesuf} {et~al.}(2014){Yesuf}, {Faber}, {Trump}, {Koo}, {Fang},
  {Liu}, {Wild}, \& {Hayward}}]{yesuf+2014}
{Yesuf}, H.~M., {Faber}, S.~M., {Trump}, J.~R., {et~al.} 2014, \apj, 792, 84

\bibitem[{{Yi} {et~al.}(1997){Yi}, {Demarque}, \& {Oemler}}]{yi+97}
{Yi}, S., {Demarque}, P., \& {Oemler}, Jr., A. 1997, \apj, 486, 201

\bibitem[{{Young} {et~al.}(2014){Young}, {Scott}, {Serra}, {Alatalo}, {Bayet},
  {Blitz}, {Bois}, {Bournaud}, {Bureau}, {Crocker}, {Cappellari}, {Davies},
  {Davis}, {de Zeeuw}, {Duc}, {Emsellem}, {Khochfar}, {Krajnovi{\'c}},
  {Kuntschner}, {McDermid}, {Morganti}, {Naab}, {Oosterloo}, {Sarzi}, \&
  {Weijmans}}]{young+2014}
{Young}, L.~M., {Scott}, N., {Serra}, P., {et~al.} 2014, \mnras, 444, 3408

\end{thebibliography}

 \end{document}